\documentclass [11pt,a4paper]{article} 
\usepackage{amsmath,amssymb,amsthm,amscd,eucal,epsf}
\usepackage{graphics}

\def\Blackboardfont{\mathbb}

\def\Z{{\Blackboardfont Z}}
\def\M{{\Blackboardfont M}}
\def\N{{\Blackboardfont N}}
\def\R{{\Blackboardfont R}}
\def\cU{{\cal U}}
\def\cP{{\cal P}}
\def\cV{{\cal V}}
\def\cS{{\cal S}}
\def\cI{{\cal I}}
\def\eref#1{(\ref{#1})}

\newtheorem{proposition}{Proposition}[section]

\newtheorem{corollary}[proposition]{Corollary}
\newtheorem{lemma}[proposition]{Lemma}

\newtheorem{example}[proposition]{Example}

\newcounter{Euro}
\begin{document}

\title{ \bf Computing the average parallelism in trace monoids\thanks{This work was partially supported by the European Community Framework
  \Roman{Euro} programme through the research network {\sc ALAPEDES}
  (``The ALgebraic Approach to Performance Evaluation of Discrete
  Event Systems'').}} 

\author{Daniel Krob\,\thanks{{\sc Liafa}, {\sc Cnrs} - Universit\'e
               Paris 7 -  Case 7014 - 2, place Jussieu - 75251 Paris Cedex 5 -
                France - \{dk,mairesse,michos\}@liafa.jussieu.fr}
        \and
        Jean Mairesse\,\footnotemark[2] \and 
        Ioannis Michos\,\footnotemark[2]}

\maketitle

\begin{abstract}
The {\em height} of a trace is the height of the corresponding
heap of pieces in Viennot's representation, or equivalently
the number of factors in its Cartier-Foata decomposition. 
Let $h(t)$ and $|t|$ stand respectively for the height and the length of a
trace $t$. We prove that the bivariate commutative series 
$\sum_t x^{h(t)}y^{|t|}$ is rational, and we give a finite 
representation of it. We use the rationality to obtain precise information on the asymptotics 
of the number of traces of a given height or length. 
%For highly symmetric trace monoids, the computations may become very
%effective. 
Then, we study the average height of a trace for various probability distributions 
on traces. For the uniform probability distribution on traces of the same length (resp. of the same height),
the asymptotic average height (resp. length) exists and is an algebraic number. 
To illustrate our results and methods, 
we consider a couple of examples: the free commutative monoid and the trace monoid whose independence graph 
is the ladder graph. 
%we consider the family of
%trace monoids whose dependence graphs are 
%{\em triangular graphs}, i.e., line graphs of the
%complete graphs. We study the
%combinatorics of this case in details.
\vspace{0.15in}

\end{abstract}

%{\bf Classification (MSC 1991)}: ??????

\section{Introduction}
Traces are used to model the occurrence of events in concurrent systems \cite{trac}.
Roughly speaking, a letter corresponds to an event and two letters commute when the
corresponding events can occur simultaneously. In this context, the two basic 
performance measures associated with a trace $t$ are its {\em length} $|t|$ (the `sequential' 
execution time) and its {\em height} $h(t)$ (the `parallel' execution time).
The ratio $|t|/h(t)$ captures in some sense the {\em amount of parallelism} 
(the {\em speedup} in \cite{CePe}).
Let $\M$ be a trace monoid. Define the generating series
\[
F=\sum_{t\in \M} x^{h(t)}y^{|t|}, \ L=\sum_{t\in \M} y^{|t|}, \ H=\sum_{t\in \M}x^{h(t)}\:.
\]
It is well known that $L$ is a rational series \cite{CaFo}.
We prove that $F$ and $H$ are also rational and we provide finite representations 
for the series. Exploiting the symmetries of the trace monoid enables to 
obtain representations of reduced dimensions.
We use the rationality to obtain precise information on the asymptotics 
of the number of traces of a given height or length. 

Then, given a trace monoid and
a measure on the traces,
we study the {\em average parallelism} in the trace monoid. 
One notion of average parallelism is obtained by considering the measure over traces induced 
by the uniform distribution over words of the same length in the free monoid. In other terms,
the probability of a trace is proportional to the number of its re\-pre\-sen\-ta\-ti\-ves in the free monoid. 
This quantity was introduced in \cite{sahe} and later studied in
\cite{BGPa,bril,BrVi98,GaMa95,SaZe}. Here we define alternative notions 
of average parallelism by considering successively the uniform distribution over 
traces of the same length, the uniform distribution over 
traces of the same height, and the uniform distribution 
over Cartier-Foata normal forms. We prove in particular that there exists 
$\lambda_{\M}$ and $\gamma_{\M}$ in $\R_+^*$ such that
\[
\frac{\sum_{t\in \M, |t|=n} h(t)}{n\cdot \#\{t\in \M,|t|=n\}} \stackrel{n\rightarrow \infty}{\longrightarrow} \lambda_{\M}, \ \  
\frac{\sum_{t\in \M, h(t)=n} |t|}{n\cdot \#\{t\in \M,h(t)=n\}} \stackrel{n\rightarrow \infty}{\longrightarrow} \gamma_{\M}\:.
\]
Furthermore, the numbers $\lambda_{\M}$ and $\gamma_{\M}$ are algebraic. 
Explicit formulas involving the series $L$ and $H$ are given for $\lambda_{\M}$ and $\gamma_{\M}$. 

The present paper is an extended version with proofs of \cite{KMMi02}.

%\vspace{0.3in}

\section{The Trace Monoid}
We start by introducing all the necessary notions from the theory of
trace monoids. The reader may refer to \cite{DiMe,trac} for further
information. 

\medskip

In the sequel, a {\em graph} is a couple $(N,A)$ where $N$ is a finite non-empty set and $A\subset N\times N$. 
Hence we consider directed graphs, allowing for self-loops but not multi-arcs. 
Such a graph is {\em non-directed} if $A$ is symmetric. 
We use without recalling it the basic terminology
of graph theory. 
Given a graph and two nodes $u$ and $v$, we write $u\rightarrow v$ if there is a path from $u$ 
to $v$. 

\medskip

Fix a finite alphabet $\Sigma$.  Let $D$ be a 
reflexive and symmetric relation on $\Sigma$, called the {\em dependence}
  relation, and let $I$ be its complement in $\Sigma \times
\Sigma$, known as the {\em independence}
or {\em commutation} relation. 

\medskip

The {\em trace monoid}, or {\em free partially commutative monoid},  ${\M}= {\M}(\Sigma,\, D)$ is
defined as the quotient of the free monoid $\Sigma^*$
by the least congruence containing
the relations $ab \sim ba$ for every $(a,b)\in I$. The elements of
${\M}$ are called {\em traces}. 
Two words are representatives of the same trace if they can be obtained one
from the other by repeatedly commuting independent adjacent letters. 

\medskip

The {\em length} of the trace $t$ is the length of any of its
representatives and is denoted by $|t|$. Note that we also use the notation $|S|=\# S$ for
the cardinal of a set $S$. 
The set of letters appearing in (any representative of)
the trace $t$ is denoted by $\mbox{alph}(t)$. 
The graphs $(\Sigma, D)$ and $(\Sigma, I)$ are called respectively 
the {\em dependence} and the {\em independence graph} of $\M$. 
Let finally $\psi$ denote the canonical projection from $\Sigma^*$
into the trace monoid $\M$. In the sequel, we most often simplify the notations by denoting a trace by 
any of its representatives, that is by identifying $w$ and $\psi(w)$. 

\begin{example}\label{ex-k4}\rm
Let $\Sigma=\{\{1,2\},\{1,3\},\{1,4\},\{2,3\},\{2,4\},\{3,4\}\}$ (the set of subsets of cardinal two
of $\{1,2,3,4\}$). Define the independence relation $I=\{(u,v) \ : \ u\cap v =\emptyset\}$. 
The dependence graph $(\Sigma,D)$
is the line graph of the complete graph $K_4$, also called 
the triangular graph $T_4$. 
\begin{figure}[htb]
\[ \epsfxsize=250pt \epsfbox{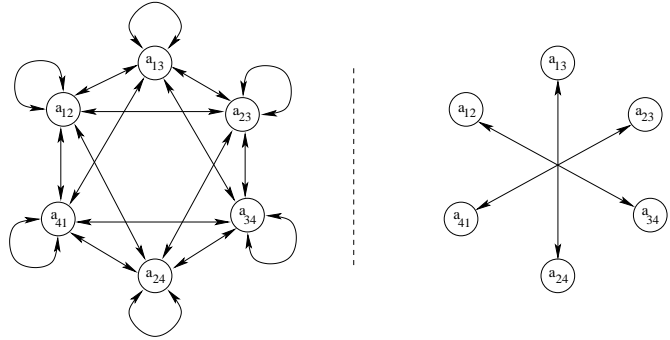} \]
\caption{The dependence graph $T_4$ (left) and its independence graph (right).}
\label{fi-t4}
\end{figure}
For notational simplicity, set $a_{ij}=\{i,j\}$. The dependence 
graph is represented on the left of Figure \ref{fi-t4} and the independence graph on the right. 
In the trace monoid ${\M}(\Sigma,D)$, we have
$\tau=a_{12}a_{34}a_{23}^2a_{14}= a_{34}a_{12}a_{23}a_{14}a_{23}$. 
\end{example}

A {\em clique} is a non-empty trace whose letters are mutually
independent. Cliques are in one-to-one correspondence with the
complete subgraphs (also called cliques in a graph theoretical context) of $(\Sigma, I)$.
We denote the set of cliques of ${\M}$ by $\mathfrak C$.

\medskip

An element $(u,v) \in {\mathfrak C} \times {\mathfrak C}$ 
is called {\em Cartier-Foata (CF-) admissible} if
for every $b \in \mbox{alph}(v)$, there exists $a \in \mbox{alph}(u)$ such that $(a,b) \in D$.
The {\em Cartier-Foata (CF) decomposition} of a trace $t$ is the uniquely
defined (see \cite[Chap. I]{CaFo}) sequence 
of cliques $(c_1 , c_2 , \dots , c_m)$ such that  
$t = c_1 c_2 \cdots c_m$, and the couple $(c_{j}, c_{j+1})$ is CF-admissible
for all $j$ in $\{1, \dots, m-1\}$.
The positive integer $m$ is called the {\em height} of $t$ and is
denoted by $h(t)$. In the visualization of traces  using
{\em heaps of pieces}, introduced by Viennot in \cite{vien},
the height corresponds precisely to the height
of the heap. 
%Also in the language of task resource models (see \cite{GaMa95}), the height is
%equal to the {\em makespan}, or execution time, of the schedule
%associated with the trace.

\begin{example}\label{ex-heapk4}\rm
Consider the trace monoid defined in Example \ref{ex-k4}. The set of
cliques is ${\mathfrak C}=
\{a, a\in \Sigma\} \cup \{ a_{12}a_{34}, a_{13}a_{24},a_{14}a_{23}\}$. 
\begin{figure}[htb]
\[ \epsfxsize=280pt \epsfbox{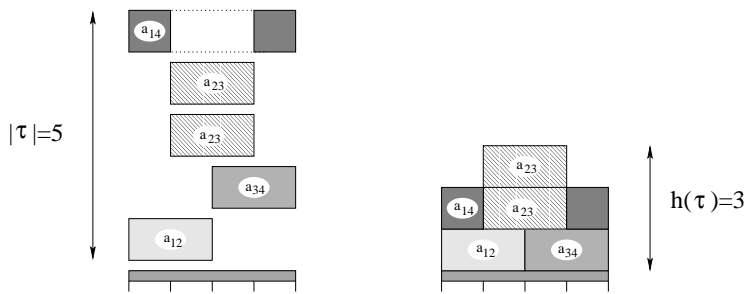} \]
\caption{Heap of pieces.}
\label{fi-heapt4}
\end{figure}
The CF decomposition of $\tau$ is $(a_{12}a_{34}, a_{14}a_{23}, a_{23})$. 
We have $|\tau|=5$  and $h(\tau)=3$. We represented the heap of pieces associated with $\tau$ 
on Figure \ref{fi-heapt4}.
\end{example}

\section{The Graph of Cliques}\label{se-gc}

We define the {\em graph of cliques} $\Gamma$ as the directed graph
with $\mathfrak C$ as its set of nodes and the set of all CF-admissible
couples as its set of arcs. 
Note that $\Gamma$ contains as a subgraph  the 
dependence graph $(\Sigma , D)$.
% (viewed as a directed graph 
%with edges being transformed into two-sided arcs).
The graph $\Gamma$ is in general complicated and looks like a
maze. 

\begin{example}\label{ex-cliqk4}\rm
We continue with the model of Examples \ref{ex-k4} and \ref{ex-heapk4}. 
\begin{figure}[htb]
\[ \epsfxsize=120pt \epsfbox{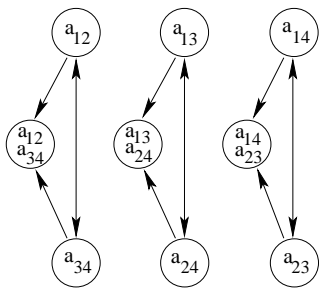} \]
\caption{The complement of the graph of cliques of $T_4$.}
\label{fi-t4cliq}
\end{figure}
For simplicity, the graph represented Figure \ref{fi-t4cliq} is the complement 
of the corresponding graph of cliques (the complement of the graph $(N,A)$ is the graph $(N,(N\times N) - A)$). 
\end{example}

\begin{lemma}\label{le-stco}
If the dependence graph 
is connected, then the corresponding graph of cliques 
is strongly connected.
\end{lemma}

\begin{proof}
Let $(\Sigma,D)$ be the dependence graph, $\mathfrak C$ the set of cliques, and
$\Gamma$ the graph of cliques. 
%For convenience, we identify a trace of length one and its unique representative.
Given $u,v\in \mathfrak C$, we want to prove that there is a path from $u$ to $v$
in $\Gamma$. 
We argue by induction on the value of $|u| + |v|$. 
If $|u| + |v| = 2$,
the result follows by the
connectivity of the dependence graph $(\Sigma,D)$. 

Now consider the case $|u| + |v| > 2$. Assume first that $|u|>1$. Let 
$a$ belong to $\mbox{alph}(u)$. Clearly $(u,a)$ is CF-admissible. 
By induction, we have $a\rightarrow v$ and we deduce that $u\rightarrow v$. 

Assume now that $|u|=1$. Then we have $|v|>1$ and let $v=v'ab, a,b\in \Sigma$.
By induction, we have $u\rightarrow v'a$. Let us prove that $v'a\rightarrow v'ab$. 
By connectivity, there exists in $(\Sigma,D)$ a path $(c_0=a,\cdots , c_k=b)$.
For $j\in \{0,\dots , k\}$, set $v_j=v'ac_j$ if $v'ac_j\in {\mathfrak C}$ and otherwise set
$v_j=wc_j$ where $w$ is the longest trace such that 
$\mbox{alph}(w)\subset \mbox{alph}(v'a)$ and $wc_j\in {\mathfrak C}$.
By construction, we obtain that $(v_0=v'a,\dots , v_k=v'ab)$ 
is a path in $\Gamma$. It completes the proof.
%define $E_j=\{x \ : \ x\in \mbox{alph}(v'), x \not\in \mbox{alph}(d_j)\}$ and let $e_j$ 
%be the clique with set of letters $E_j$.
%We choose the path to be of minimal length $l$.
%By minimality, two non-successive
%letters in the path commute. It implies that 
%we have $c_i\rightarrow \psi(ab)$ for $i\in \{1,\dots ,l-1\}$.
%Let us fix $i$ and let $(d_0=c_i,\dots, d_k=\psi(ab))$ be a path in $\Gamma$. 
%Let $v'$ be the trace such that $v=v'\psi(ab)$. 
%For $j\in \{0,\dots , k\}$,
%define $E_j=\{x \ : \ x\in \mbox{alph}(v'), x \not\in \mbox{alph}(d_j)\}$ and let $e_j$ 
%be the clique with set of letters $E_j$. We obtain by construction that
%$(d_0e_0,\dots, d_ke_k)$ is a path in $\Gamma$. Since $d_ke_k=v$, we have proved that $d_0e_0\rightarrow v$.
%Remarking that $|d_0e_0|<|v|$, we obtain by induction that $u\rightarrow d_0e_0$. It complete the proof.
\end{proof}

The above lemma can be restated as follows:
given two cliques $u$ and $v$ there exists at
least one trace in $\M$ such that the first factor in its CF-decomposition
is $u$ and the last one is $v$.

%{\em Example 2.3} Consider the dependence graph $C_4$ (the cycle of
%length $4$). Its digraph of cliques ${\Gamma}(C_4)$ is the digraph of
%order $6$ represented in Figure $1$.
%\paragraph{Front divisors of the digraph of cliques} $ $\\ 

\medskip

We now use a standard reduction technique for 
multi-graphs (see \cite[Chap. 4]{CDSa} or \cite[Chap. 5]{gods}).
We partition the nodes of $\Gamma$ based on their set of direct successors.
An {\em equitable partition} of $\mathfrak
C$ is a partition 
$\pi = \{ {\mathfrak C_{1}}, \dots, {\mathfrak C_{s}} \}$
with the property that for all $i$ and $j$ the number $a_{ij}$ of direct successors
that a node in $\mathfrak C_{i}$ has in $\mathfrak C_{j}$ is independent
of the choice of the node in $\mathfrak C_{i}$. 
The $s \times s$ matrix $A_{\pi}=(a_{ij})_{i, j}$ is called  
the {\em coloration matrix} corresponding to $\pi$.
%The valued graph with $\pi$ as the set of
%nodes and with an arc from $\mathfrak C_{i}$ to $\mathfrak C_{j}$ valued by
%$a_{ij}$ if $a_{ij}\neq 0$, is called the {\em front divisor} corresponding to $\pi$
%and will be denoted by ${\Gamma}_{\pi}$. 
In the case of the partition $\{\{c\},c\in{\mathfrak C}\}$, 
%the front divisor is $\Gamma$ itself and 
the coloration matrix is the {\em adjacency matrix} of $\Gamma$.
% that we denote by $A$. 

%Let us fix a total order on $\Sigma$ and let us denote by $\leq $ the induced
%lexicographical total order
%on $\Sigma^n$ for $n\in\N$. On $\Sigma^*$, we define
%a total order, still denoted by $\leq$, as follows:
%$u\leq v$ if $|u|<|v|$ or $|u|=|v|, u\leq v$. 
%By convention, we always assume that the ordering of the elements 
%${\mathfrak C_{1}}, \dots, {\mathfrak C_{s}},$ of a partition satisfies 
%$c_1 \leq c_2 \leq \cdots \leq c_s$, where $c_i$ is the minimal representative of 
%an element of ${\mathfrak C_{i}}$. 

\begin{example}\label{ex-colk4}\rm
We keep studying the model of Examples \ref{ex-k4}, \ref{ex-heapk4} and \ref{ex-cliqk4}. 
Consider the partition $\pi$ of $\mathfrak C$ defined by 
\[
{\mathfrak C}_1=\{a_{12},a_{13}, a_{14} \},\
{\mathfrak C}_2=\{a_{23},a_{24}, a_{34} \},\ 
{\mathfrak C}_3=\{ a_{12}a_{34}, a_{13}a_{24},a_{14}a_{23}\}\:.
\]
It is easily checked
that the partition is equitable. %The adjacency matrix $A$ and the coloration
%matrix $A_{\pi}$ are 
%\[
%A=\left(\begin{array}{ccccccccc}
% 1 & 1 & 1 & 1 & 1 & 0 & 0 & 1 & 1 \\ 
% 1 & 1 & 1 & 1 & 0 & 1 & 1 & 0 & 1 \\ 
% 1 & 1 & 1 & 0 & 1 & 1 & 1 & 1 & 0 \\ 
% 1 & 1 & 0 & 1 & 1 & 1 & 1 & 1 & 0 \\ 
% 1 & 0 & 1 & 1 & 1 & 1 & 1 & 0 & 1 \\ 
% 0  & 1 & 1 & 1 & 1 & 1 & 0 & 1 & 1 \\ 
% 1 & 1 & 1 & 1 & 1 & 1 & 1 & 1 &  1\\ 
% 1 & 1 & 1 & 1 & 1 & 1 & 1 & 1 &  1\\ 
% 1 & 1 & 1 & 1 & 1 & 1 & 1 & 1 & 1
%\end{array}\right), \ \ A_{\pi}= \left( \begin{array}{ccc}  3 & 2 & 2 \\ 
%                                                            2 & 3 & 2  \\
%                                                            3 & 3 & 3
%                          \end{array}                        \right)\:. 
The corresponding coloration matrix is
\[ A_{\pi}= \left( \begin{array}{ccc}  3 & 2 & 2 \\ 
                                                            2 & 3 & 2  \\
                                                            3 & 3 & 3
                          \end{array}                        \right)\:. 
\]
%The matrix is arranged according to the above convention with the following
%order on $\Sigma$: $a_{12}<a_{13}<a_{14}<a_{23}<a_{24}<a_{34}$. 
\end{example}

A natural family of equitable partitions is the one induced by the non-trivial 
subgroups of the full automorphism group of $\Gamma$. 
Given such a group $G$, the cells of the corresponding partition $\pi_G$ 
are the orbits into which $\mathfrak C$ is partitioned by $G$.
The corresponding coloration matrix is denoted by $A_{G}$.

An automorphism of $(\Sigma,D)$ induces an automorphism of $\Gamma$. 
Indeed, consider an automorphism $\phi$ of $(\Sigma, D)$. 
The map $\phi: \Sigma \rightarrow \Sigma $ can be extended into a map 
$\phi': {\mathfrak C} \rightarrow {\mathfrak C}$ as follows. 
Given $c=u_1\cdots u_k\in {\mathfrak C}$ with $|u_i|=1$ for all $i$, set
$\phi'(c)=\phi(u_1)\cdots \phi(u_k)$. 
Note that the definition is unambiguous since the letters $\phi(u_i)$ commute.  
It is immediate that $\phi'$ is an automorphism of $\Gamma$. 

Due to the complex structure of $\Gamma$, finding its
automorphisms is in general difficult. 
Finding the automorphisms of $(\Sigma,D)$ is often an easier task. 
This simple observation
allows us to focus on the automorphism groups of $(\Sigma, D)$
and to consider their action on the nodes of ${\Gamma}$. 
When $(\Sigma, D)$ has a great amount of symmetries,
the corresponding reduction can be very important (see 
section \ref{sse-hsg}). 

\medskip

Below we need to consider  equitable partitions such that 
all the cliques in the same cell have a common length. This requirement 
is always satisfied for the equitable partitions associated with automorphism groups.

\begin{example}\label{ex-colok4}\rm
The model is the one of Examples \ref{ex-k4}, \ref{ex-heapk4}, \ref{ex-cliqk4}, and \ref{ex-colk4}. 
The symmetric group ${\mathfrak S}_4$ of degree 4 is a non-trivial group of automorphisms of $(\Sigma,D)$. 
It is of index $2$ in the full automorphism group $G$ of $(\Sigma,D)$.
% (see \cite[Lemma 3]{higm}).
The partition of ${\mathfrak C}$ induced by ${\mathfrak S}_4$ (or by $G$) 
is  ${\mathfrak C}_1=\{a, a\in \Sigma\}$ 
and ${\mathfrak C}_2=\{ a_{12}a_{34}, a_{13}a_{24},a_{14}a_{23}\}$.
The coloration matrix is given by
\[
A_{{\mathfrak S}_4}= \left( \begin{array}{cc}  5 & 2 \\ 6 & 3
                \end{array}                        \right)\:. 
\]
\end{example}

\section{Height and Length Generating Function}\label{se-hlgf}

Let $F \in \N[[x,y]]$ be the {\em  height and length generating function}
defined by
\[ F(x, y) \: = \: \sum_{t \in {\M}} x^{h(t)}y^{|t|} 
           \: = \: \sum_{k,\,l\ \in \N} \, f_{k,l} \, x^{k} y^{l}\:,\]
where $x$ and $y$ are commuting indeterminate and $f_{k,l}$ is the number 
of traces of height $k$ and length $l$. Set $H(x) = F(x,1)$ and $L(y) = F(1,y)$. 
Then $H(x)$ and $L(y)$ are respectively the generating functions of the height and of the length.
The {\em M\"{o}bius polynomial} $\mu(\Sigma,I)$ of the graph $(\Sigma,I)$ is defined by 
\begin{equation}\label{eq-mobius}
\mu(\Sigma,I) = 1+ \sum_{u \in {\mathfrak C}} (-1)^{|u|} y^{|u|}\:.
\end{equation}

It is well known \cite[Chap. II]{CaFo} that $L(y)$
is equal to the inverse of the M\"{o}bius polynomial, i.e. $L(y)= \mu(\Sigma,I)^{-1}$. In particular,
it is a rational series. 
%\begin{equation}\label{eq-mobius}
%L(y)= \mu(\Sigma,I)^{-1}\:.
%\end{equation}
%For the height and length generating function, we prove the following result.

\begin{proposition}\label{th-A}
Let $\M=\M(\Sigma,D)$ be a trace monoid and let $\mathfrak C$ be the set of
cliques of $(\Sigma,I)$. 
Define the matrix $A(x,y)\in {\N}[x,y]^{\mathfrak C \times \mathfrak C}$ by setting
$A(x,y)_{i,j} = xy^{|i|}$ if $(i,j)$ is CF-admissible and 0 otherwise. Define also
$u = (1,\dots , 1) \in {\N}[x,y]^{1
  \times \mathfrak C}$ and $v(x,y) = (xy^{|i|})_i \in {\N}[x,y]^{\mathfrak C \times 1}$. The height and length 
generating function is then given by
\begin{equation}\label{eq-F}
 F - 1 \, = \, \sum_{n \in \N} u A(x,y)^{n} v(x,y)
 \, = \, u \left( I - A(x,y) \right)^{-1}v(x,y)\,,
\end{equation}
where $1$ is the identity of $\N[[x,y]]$ and $I$ is the $\mathfrak C \times \mathfrak C$ identity matrix. 
\end{proposition}

Proposition \ref{th-A} states that $F(x,y)$ is a rational series of $\N[[x,y]]$ and that
$(u,A(x,y),v(x,y))$ is a finite representation of it. 

\begin{corollary}\label{co-A}
The series $L$ and $H$ are rational,
and we have $L=1+u \left( I - A(1,y) \right)^{-1}v(1,y)$ and $H=1+u \left( I - xA(1,1) \right)^{-1}xv(1,1)$. 
\end{corollary}

Proposition \ref{th-A} and Corollary \ref{co-A}, although easy to prove, do not seem 
to appear in the literature. In the case of the length generating series, the rationality is not new but Corollary
\ref{co-A} provides a new formula for $L$. 

There exist related results in the context of directed animals. 
Indeed there is a bijection between directed animal of width $k$ on a 2d triangular lattice 
and traces in the monoid $\M(\Sigma,D)$ with $\Sigma=\{a_1,\dots,a_k\}$ and $D=\{(a_i,a_j), |i-j|\leq 1\}$. 
The precise asymptotics for such directed animals are derived in 
\cite{HaNa,NDVa} with the same method as in the proof of Proposition \ref{th-A}. 
More generally, the method of proof of Proposition \ref{th-A} can be viewed as an instance
of the transfer matrix method \cite[Chap. 4.7]{stan}. 

In the context of trace monoids, 
the idea of working with the alphabet of cliques ${\mathfrak C}$ to study the height function 
appeared in \cite{CePe} and was later used in \cite{GaMa98}. 

\medskip

Let ${\pi}= \{ {\mathfrak C_{1}}, \dots, {\mathfrak C_{s}} \}$ be an
equitable partition of $\mathfrak C$ 
such that all the cliques in $\mathfrak C_i$ have a common length $l_i$. 
Let 
$A_{\pi}=(a_{ij})_{ij}\in \N^{s\times s}$ be the coloration matrix.
Define the matrix $A_{\pi}(x,y)\in {\N}[x,y]^{s\times s}$ by 
$A_{\pi}(x,y) = (a_{ij}xy^{\,l_i})_{i,j}$.
Define 
$u_{\pi} = (|{\mathfrak C}_i|)_i \in {\N}[x,y]^{1
  \times s}$ and $v_{\pi}(x,y) = (xy^{l_i})_i \in {\N}[x,y]^{s \times 1}$. 
Then formula \eref{eq-F} holds when replacing $u,A(x,y)$, and $v(x,y),$ by $u_{\pi},A_{\pi}(x,y)$, and 
$v_{\pi}(x,y)$.
The proof is similar to the one below. 

\begin{proof}[Proof of Proposition \ref{th-A}]
As recalled above, with each trace is associated its unique CF decomposition. 
We associate with a path $p$ in $\Gamma$ the sequence of its nodes
$(c_1,\dots, c_k)$.
By construction, the CF decomposition of the trace $t=c_1\cdots c_k$ 
is precisely $(c_1,\dots,c_k)$. In other words,
the CF decompositions of traces are in one-to-one correspondence 
with the paths in $\Gamma$. 
The contribution of the trace $t$ to the series $F$ is $x^{h(t)}y^{|t|}$. The weight of 
the path $p$ in the weighted automaton $(u,A(x,y),v(x,y))$ is 
\[
u_{c_1}(\prod_{i=1}^{k-1}A(x,y)_{c_ic_{i+1}}) v(x,y)_{c_k}= (\prod_{i=1}^{k-1}xy^{|c_i|})xy^{|c_k|}=x^{h(t)}y^{|t|}\:.
\]
This completes the proof of the result. 
\end{proof}

It is easily checked that the series $F(x,y)$ is not recognizable in general. We recall that 
$F= \sum_{k,l} f_{k,l} x^ky^l$ is a {\em recognizable} series  of $\N[[x,y]]$ if there exists 
$K\in \N^*, \alpha \in \N^{1\times K}, \mu(x)\in \N^{K\times K}, \mu(y)\in \N^{K\times K}$, and
$\beta \in 
\N^{K\times 1}$, such that $f_{k,l}=\alpha\mu(x)^k\mu(y)^l\beta$ for all $k$ and $l$. 

\begin{example}\label{ex-thk4}\rm
We persevere with the model of Examples \ref{ex-k4}, \ref{ex-heapk4}, \ref{ex-cliqk4}, 
\ref{ex-colk4}, and \ref{ex-colok4}. 
The height and length generating function is given by
\begin{eqnarray}\label{eq-genfun}
F & = &   1+ \left(\begin{array}{cc} 6 & 3 \end{array}\right) 
\left(\begin{array}{cc} 1-5xy & -2xy \\ -6xy^2 & 1- 3xy^2 \end{array}\right)^{-1}
\left(\begin{array}{c} xy \\ xy^2 \end{array}\right) \nonumber \\
& = & \frac{1+xy}{1-5xy-3xy^2+3x^2y^3}\:.
\end{eqnarray}
Setting $x=1$, we  check that the length generating function is the inverse 
of the M\"{o}bius polynomial, i.e. $L=(1-6y+3y^2)^{-1}$. 
Setting $y=1$, we obtain the height generating function $H=(1+x)(1-8x+3x^2)^{-1}$. 
The Taylor expansion of the series $F$ around 0 is 
\begin{eqnarray*}
F&=&1 + 6xy + 3xy^2 + 30x^2y^2 + 30x^2y^3 +150 x^3y^3 + 9x^2y^4 + 222x^3y^4 + \\
&& 750 x^4y^4+ 
126x^3y^5+ 1470 x^4y^5+ \cdots+ 71910x^6y^8+\cdots\:.
\end{eqnarray*}
For instance, there are 126 traces of length 5 and height 3, or 71910 traces of length 8 and height 6.
\end{example}

We now use Proposition \ref{th-A} to provide some precise results on the asymptotics 
of the number of traces of a given length or height. 

\medskip

Given a complex function analytic at the origin, a
{\em singularity} is a point where the function ceases to be complex-differentiable. 
A {\em dominant} singularity  is a singularity of minimal modulus.
Throughout the paper, given a series $S\in \N[[x]]$, we set $S=\sum_n (S|n)x^n$. 
When applicable, we denote the modulus
of the dominant singularities of $S$ (viewed as a function) by $\rho_S$. 
Classically, see \cite{BeRe,FlSe,wilf}, the asymptotic growth rate of $(S|n)$ is linked to the
values of the dominant singularities.

\begin{lemma}\label{le-rho=1}
We have $\rho_L=1$ or $\rho_H=1$ if and only if $\M(\Sigma,D)$ is the free commutative monoid over 
$\Sigma$. 
\end{lemma}

\begin{proof}
We have $\limsup_n (L|n)^{1/n} = 1/\rho_L$, and 
$\limsup_n (H|n)^{1/n} = 1/\rho_H$ (the `exponential growth formula').
It implies that $\rho_L\leq 1$ and $\rho_H\leq 1$. 

Assume there exists $(a,b)\in D$ with $a\neq b$. 
Then all the traces $t_1\cdots t_n$ with $t_i\in \{a,b\}$ are of length $n$ and height 
$n$. It implies that $(L|n)\geq 2^n$ and that $(H|n)\geq 2^n$.
It implies in turn that $\rho_L\leq 1/2$ and $\rho_H\leq 1/2$. 

Assume now that $\M(\Sigma,D)$ is the free commutative monoid. 
By direct computation or using the results from section \ref{sse-fcm}, we get 
$(L|n)\sim n^{|\Sigma|-1}$
and $(H|n)\sim n^{|\Sigma|-1}$. It implies 
that $\rho_L=1$ and $\rho_H=1$. 
\end{proof}

\begin{proposition}\label{le-clique}
Let $(\Sigma,D)$ be a connected dependence graph. 
Then $L$ and $H$ have a unique dominant singularity which is 
positive real and of order 1. 
\end{proposition}

It follows (see \cite{BeRe,FlSe,wilf}) that when $(\Sigma,D)$ is connected, we have 
$(L|n) \sim \alpha_L\rho_L^{-n}$ and 
$(H|n)\sim \alpha_H\rho_H^{-n}$, with $\alpha_L=\rho_L^{-1}\cdot [L(y)(\rho_L-y)]_{|y=\rho_L}$ and
$\alpha_H=\rho_H^{-1}\cdot [H(x)(\rho_H-x)]_{|x=\rho_H}$. 

\medskip

The proof of Proposition \ref{le-clique} is based on the representation given in Proposition \ref{th-A}. 
For convenience reasons, the proof  is included in the proof
of Proposition \ref{pr-lambda2} and given in Appendix. 

\begin{proposition}\label{le-clique2}
Let $(\Sigma,D)$ be a non-connected dependence graph. Let $(\Sigma_s,D_s)_{s\in S}$ be its partition
into maximal connected subgraphs. Denote by $L_s,H_s$, the 
corresponding length and height generating functions. Then one has:

\noindent 
1) the series $L$
has a unique dominant singularity equal to $\rho_L=\min_s \rho_{L_s}$, 
and whose order is $\#\{s ,  \rho_{L_s}=\rho_L\}$;

\noindent
2) the series $H$ has a unique dominant singularity equal to $\rho_H=\prod_s \rho_{H_s}$.
Its order is $|\Sigma|$ if $\M(\Sigma,D)$ is the free commutative monoid, and
$1+\#\{s ,  |\Sigma_s|=1\}$ otherwise.
\end{proposition}

Let $k_L$ and $k_H$ denote the respective orders of  $\rho_L$ in $L$ and $\rho_H$ in $H$. 
It follows from the above Proposition (see \cite{BeRe,FlSe,wilf}) that we have 
$(L|n) \sim  \alpha_L n^{k_L-1}\rho_L^{-n}$, and
$(H|n)\sim  \alpha_H n^{k_H-1}\rho_H^{-n}$ with 
$\alpha_L=(\rho_L^{-k_L}/(k_L-1)!)\cdot [L(y)(\rho_L-y)^{k_L}]_{|y=\rho_L}$
and $\alpha_H=(\rho_H^{-k_H}/(k_H-1)!)\cdot [H(x)(\rho_H-x)^{k_H}]_{|x=\rho_H}$.  

\begin{proof}
We have $L=\prod_s L_s= \prod_s \mu(\Sigma_s,I_s)^{-1}$ where $\mu(.)$ is defined
in \eref{eq-mobius}. It implies directly the result on $\rho_L$. 

\medskip 

Consider now the height generating function. We  prove the result by induction 
on $|S|$. 
Assume first that $\# S=2$ and set $S=\{1,2\}$. 
We have $H=\sum_{i,j} (H_1|i)(H_2|j)x^{\max(i,j)}$. It implies that
\begin{equation}\label{eq-hh1h2}
(H|n)= (H_1|n) \sum_{i=0}^n (H_2|i) + (H_2|n)\sum_{i=0}^n (H_1|i) - (H_1|n)(H_2|n)\:.
\end{equation}
Applying Proposition \ref{le-clique}, we obtain
$(H_1|n)= a_n \rho_{H_1}^{-n}$, with $\lim_n a_n=a\in \R_+^*$, and 
$(H_2|n)= b_n \rho_{H_2}^{-n}$, with $\lim_n b_n=b\in \R_+^*$. 

\medskip

Consider first the case $\rho_{H_1} < 1$ and $\rho_{H_2}<1$. 
We have 
\begin{eqnarray*}
(H_1|n) \sum_{i=0}^n (H_2|i) & = & a_n \rho_{H_1}^{-n} (\sum_{i=0}^n b_i \rho_{H_2}^{-i}\ ) \\
& = & a_n b \rho_{H_1}^{-n}\rho_{H_2}^{-n}(\sum_{i=0}^n (b_{n-i}/b) \rho_{H_2}^{i}\ ) \\
& \sim  & ab(1-\rho_{H_2})^{-1} (\rho_{H_1}\rho_{H_2})^{-n} .
\end{eqnarray*}
The same type of identity also holds for the second term in \eref{eq-hh1h2}.
Going back to \eref{eq-hh1h2}, we then obtain
\[
(H|n) \sim ab( (1-\rho_{H_1})^{-1} + (1-\rho_{H_2})^{-1} -1 ) (\rho_{H_1}\rho_{H_2})^{-n}\:.
\]
Hence we have $\rho_H=\rho_{H_1}\rho_{H_2}$ and the order of $\rho_H$ in $H$ is 1. 

\medskip

We consider now the case $\rho_{H_1}=1$ and $\rho_{H_2}=1$. By Lemma \ref{le-rho=1}, 
we get that $\M(\Sigma,D)$ is the free 
commutative monoid over two letters.
Applying \eref{eq-hh1h2}, we get that $(H|n)=(2n+1)$. Hence we have $\rho_H=1$ and the order of $\rho_H$
in $H$ is $2$. 

By symmetry, the last case to consider is $\rho_{H_1} < 1$ and $\rho_{H_2}=1$. 
By Proposition \ref{le-clique}, we have $(H_1|n)\sim a \rho_{H_1}^{-n}$. We also have $(H_2|n)=1$. 
Simplifying \eref{eq-hh1h2}, we obtain that $(H|n)\sim a n \rho_{H_1}^{-n}$.
It implies that $\rho_H=\rho_{H_1}$ and that the order of $\rho_H$ in $H$ is 2. 

Consider now the case $\# S>2$. 
Let $(\Sigma_1,D_1)$ and $(\Sigma_2,D_2)$ be a partition of $(\Sigma,D)$
in two subgraphs such that $(\Sigma_1,D_1)$ is connected. The induction hypothesis 
applies to $(\Sigma_2,D_2)$ and the proof follows exactly the same steps as above 
\end{proof}

The results on $L$ in Proposition \ref{le-clique} and Proposition \ref{le-clique2} can be restated as results on 
the smallest root of the M\"obius polynomial of a non-directed graph. 
They improve on a recent result by Goldwurm and Santini \cite{GoSa} stating that
the M\"obius polynomial has a unique and positive real root of smallest modulus. 
Our proof of Proposition \ref{pr-lambda2} follows several of the steps of \cite{GoSa}. 
One central difference is that
we work with Cartier-Foata representatives instead of minimal lexicographic representatives. 
Proving the strengthened statements while working with the latter does not appear to be easy. 

\medskip

A {\em matching} in a (non-directed) graph is a subset of arcs with no common nodes. 
The {\em matching polynomial} of a graph is equal to $\sum_k (-1)^k m_k y^k$, where $m_k$ is the
number of matchings of $k$ arcs. Hence, 
the matching polynomial of a graph $G$ is equal to the M\"obius polynomial of the complement of the line
graph of $G$.
Matching polynomials have been studied quite extensively. It is known for instance that 
all the roots of a matching polynomial are real \cite{gods81,GoGu}. 
It implies that the same is true for the M\"obius polynomial of a graph which is the complement of a line graph. 
For a general graph, the result is not true and one has to settle for the weaker results
in Proposition \ref{le-clique} and Proposition \ref{le-clique2}. Consider for instance 
the graph with nodes $\{a,b,c,d\}$ and arcs
$\{(a,b),(b,a),(a,c),(c,a),(b,c),(c,b)\}$. 
It is the smallest graph which is not the complement of a line graph.
Its M\"obius polynomial is $\mu=1-4y+3y^2-y^3$, which has two non-real roots.

\section{Asymptotic Average Height}\label{se-aah}

We want to address questions such as: what is the amount of `parallelism'
in a trace monoid? Given several dependence graphs over the same alphabet, 
which one is the `most parallel'? To give a precise meaning to these questions, we define
the following performance measures. Let ${\M}_{\,n}$ denote the set of traces of length $n$
of the trace monoid $\M$. 
We equip ${\M}_{\,n}$ with a probability distribution $P_n$ and
we compute the 
corresponding average height 
\[
E_n[h]=\sum_{t\in {\M}_n } P_n\{t\} h(t)\:.
\]
Assuming the limit
exists, we call $\lim_n E_n[h]/n$ the {\em (asymptotic) average height}. 
Obviously this quantity belongs to $[C^{-1},1]$, where $C$ is the maximal length of a clique. 
Clearly
the relevance of the
average height as a measure of the parallelism in the trace monoid depends
on the relevance of the chosen family of probability measures. This may vary depending on the 
application context. 
A very common choice is to consider uniform probabilities. 
It is the natural solution in the absence of precise 
information on the structure of the traces to be dealt with. 
Let us consider different instances of uniform probabilities over traces.

\subsection{Uniform probability on words}
Let $\mu_n$ be the uniform probability 
distribution over $\Sigma^n$ which is defined by setting
$\mu_n\{u\}=1/|\Sigma|^n$, for every $u\in \Sigma^n$. We set
$P_n = \mu_n\circ \psi^{-1}$, i.e. $P_n\{t\}=\mu_n\{ w \ : \ \psi(w)=t\}$. 
The limit below exists:
\begin{equation}
\label{eq-lambda}
\lambda_{*} =\lambda_{*}(\Sigma,D)= 
\lim_n \frac{E_n[h]}{n} = \lim_n \frac{\sum_{w\in \Sigma^n} h(\psi(w))}{n|\Sigma|^n} \:.
\end{equation}
This is proved using Markovian arguments in \cite{sahe}.
The existence of $\lambda_{*}$ can also be proved using sub-additive arguments. 
More precisely, it is shown in  \cite{GaMa95} that $h(\psi(.))$ is recognized by an automaton
with multiplicities over the $(\max,+)$ semiring, which provides a different proof 
of the existence of $\lambda_{*}$. 
In fact a stronger result holds. Consider a probability space $(\Omega,{\cal F},P)$. 
Let $(x_n)_{n\in \N^*}$ be a sequence of independent 
random variables valued in $\Sigma$ and 
uniformly distributed: $P\{x_n=u\}=1/|\Sigma|, u\in \Sigma$. 
The probability distribution of $(x_1\cdots x_n)$ is then the uniform distribution over $\Sigma^n$. 
It is proved in \cite{sahe,GaMa95} that 
\begin{equation}\label{eq-king}
P \{ \ \lim_n \frac{h(\psi(x_1\cdots x_n))}{n} = \lambda_{*} \ \} =1\:.
\end{equation}

Except
for small trace monoids, $\lambda_{*}$ is neither rational, nor algebraic. 
The problem of approximating $\lambda_{*}$ is 
NP-hard \cite{BGT}. Non-elementary bounds are proposed in
\cite{BrVi98}. Exact computations for simple trace monoids are proposed 
in \cite{bril,SaZe}. A software package named {\sc Ers} \cite{ers} enables
to simulate and compute bounds for $\lambda_*$. 

%\medskip

%Assume that $(\Sigma,D)$ is {\em non-connected} and let 
%$(\Sigma_s,D_s), s\in S,$ be the maximal strongly connected subgraphs of $(\Sigma,D)$. 
%%Let $\lambda_{*}(\Sigma,D), \lambda_{*}(\Sigma_1,D_1), \dots , \lambda_{*}(\Sigma_k,D_k)$, 
%%be the respective average heights of
%%$(\Sigma,D),(\Sigma_1,D_1),\dots ,(\Sigma_k,D_k)$. 
%Then we have
%\begin{equation}\label{eq-decomp}
%\lambda_{*}(\Sigma,D) = \max_{s\in S} \ ( \ \frac{|\Sigma_s|}{|\Sigma|} \lambda_{*}(\Sigma_s,D_s) \ ) \:.
%\end{equation}
%This formula can be easily obtained using \eref{eq-king} and the Strong Law of Large Numbers (see
%Theorem 5.7 in \cite{sahe} for a detailed argument). 
%Let us justify this formula. 
%Let $e$ denote the empty word of the free monoid $\Sigma^*$ (resp. $\Sigma_i^*$). 
%Given $x\in \Sigma$, we define $x^{(j)}=x$ if $x\in \Sigma_j$ and $x^{(j)}=e$ otherwise. 
%Applying \eref{eq-king}, we have
%\[
%\lim_{n} \frac{h(\psi(x_1^{(j)}\cdots x_n^{(j)}))}{|x_1^{(j)}\cdots x_n^{(j)}|} \ = \ \lambda_{*}_j \:,
%\]
%almost surely. Furthermore, by the Strong Law of Large Numbers, we have 
%\[
%\lim_n \frac{|x_1^{(j)}\cdots x_n^{(j)}|}{n} \ = \ \frac{1}{n} (\sum_{i=1}^n {\bf 1}_{\{x_i\in \Sigma_j\}} )= |\Sigma_j|/|\Sigma|\:,
%\]
%almost surely. 
%Remarking that $h(\psi(x_1\cdots x_n))= \max_j h(\psi(x_1^{(j)}\cdots x_n^{(j)}))$, 
%we deduce formula \eref{eq-decomp}. 

\subsection{Uniform probability on traces}\label{sse-upt}

A natural counterpart of the above case consists in considering the 
uniform probability distribution over ${\M}_{\,n}$, i.e.
$Q_n\{t\} = 1/|{\M}_{\,n}|$ for every $t \in {\M}_{\,n}$. 
Assuming existence, we define the limit
\begin{equation}
\label{eq-lambda2}
\lambda_{\M} = \lambda_{\M}(\Sigma,D)=\lim_n \frac{E_n[h]}{n} = \lim_n \frac{\sum_{t\in \M_n} h(t)}{n|{\M}_{\,n}|}=
\lim_n \frac{\sum_{m \in \N} m f_{m,n}} {\sum_{m \in \N} nf_{m,n}}\:.    
\end{equation}

\medskip

Dually, let ${}_m \M$ be the set of traces of height $m$, and let
$\tilde{Q}_m$ be the uniform probability measure on ${}_m \M$, i.e.
$\tilde{Q}_m\{t\}= 1/|{}_m \M|$ for every $t\in {}_m \M$. The average length of a trace in ${}_m \M$ is equal to 
$E_m[l]=\sum_{t \in {}_m \M} \tilde{Q}_m\{t\} |t| $.
Assuming existence, we define the limit
\begin{equation}
\label{eq-gamma}
\gamma_{\M} = \gamma_{\M}(\Sigma,D)= \lim_m  \frac{E_m[l]}{m}= 
\lim_m \frac{\sum_{t\in {}_m \M} |t|}{m|{}_m \M |}=
\lim_m \frac{\sum_{n \in \N} n f_{m,n}} {\sum_{n \in \N} mf_{m,n}}\:.    
\end{equation}
The quantity $\gamma_{\M}$ is an {\em (asymptotic) average length}.
The analog of $\lambda_{*}$ and $\lambda_{\M}$ is then the quantity 
$\gamma_{\M}^{-1}$.

\begin{proposition}\label{pr-lambda2}
The limits $\lambda_{\M}$ in \eref{eq-lambda2} and $\gamma_{\M}$ in \eref{eq-gamma} exist. Furthermore,
$\lambda_{\M}$ and $\gamma_{\M}$
are algebraic numbers.
\end{proposition}

The proof, based on Proposition \ref{th-A}, is rather long and we postponed it to the Appendix. 
In fact, the proof of Proposition \ref{pr-lambda2} provides a formula for $\lambda_{\M}$ and $\gamma_{\M}$.
Define $G=(\partial F/\partial x)(1,y)$ and $\tilde{G}=(\partial F/\partial y)(x,1)$.
%Let $\rho$ and $\tilde{\rho}$ be respectively the unique and positive real (see the Appendix) 
%dominant singularity of $L$ and $H$ and let $k$ and $\tilde{k}$ be their corresponding order. 
Then, with the notations of section \ref{se-hlgf}, we have
\begin{equation}\label{eq-complex0}
\lambda_{\M} = \frac{ [G(y)(\rho_L-y)^{k_L+1}]_{|y=\rho_L}}{k_L\rho_L [L(y)(\rho_L-y)^{k_L}]_{|y=\rho_L}}, \ 
\gamma_{\M} = \frac{ [\tilde{G}(x)(\rho_H-x)^{k_H+1}]_{|x=\rho_H}}{k_H\rho_H
[H(x)(\rho_H-x)^{k_H}]_{|x=\rho_H}}\:.
\end{equation}

\subsection{Uniform probability on CF decompositions}\label{sse-cf}

In this section, we use some basic results on Markov chains, for details 
see for instance \cite{brem99,revu,sene}. 
Let $A\in \{0,1\}^{{\mathfrak C}\times {\mathfrak C}}$ be the adjacency matrix of $\Gamma$. 
We associate with $A=(a_{ij})_{i,j}$, the Markovian matrix 
\begin{equation}\label{eq-markov}
\widehat{A}=  (\widehat{a}_{ij})_{i,j}, \  
\widehat{a}_{ij} = a_{ij}(\sum_k a_{ik})^{-1}\:.
\end{equation}
We define the vector $\vec{1} \in \R^{1\times {\mathfrak C}}$ by 
$\vec{1}_i=1/|{\mathfrak C}|$ for all $i$.
We define the probability measure $R_m$ on ${}_m \M$ as follows: 
for a trace 
$t\in  {}_m \M$ with Cartier-Foata decomposition $(c_1,\dots , c_m)$, 
we set $R_m\{t\} = \vec{1}_{c_1} \widehat{a}_{c_1c_2}\cdots \widehat{a}_{c_{m-1}c_m}$. 

An interpretation for the family $(R_m)_m$ 
is as follows.
Consider a Markov chain $(X_n)_n$ on the state space ${\mathfrak C}$ 
with transition matrix $\widehat{A}$ and with initial distribution 
$\vec{1}$. Then $R_m\{t\}=P\{X_1\cdots X_m=t\}$. Equivalently, 
given a trace $t$ of height $m$, we get a trace $t'$ of height $m+1$ by
picking at random and uniformly an admissible clique $c$ and by setting $t'=tc$. 
This can be loosely described as a `uniform probability on CF decompositions'.

The average length of a trace in ${}_m \M$ is equal to 
$E_m[l]=\sum_{t \in {}_m \M} R_m\{t\} |t|$.
Assuming existence, the analog of $\lambda_{*}, \lambda_{\M}$ or $\gamma_{\M}^{-1}$ is then the 
{\em (asymptotic) average height} 
%By the ergodic theorem for Markov chains (Theorem 4.6 in \cite{sene}), 
%the quantity $(\lim_m E_m[l]/m)$ exists. 
%%We call $\lambda_{\mathrm{cf}}^{-1}$ the {\em (asymptotic) average length}.
%The analog of $\lambda_{*}, \lambda_{\M}$ or $\gamma_{\M}^{-1}$ is then the 
%{\em (asymptotic) average height} 
\begin{equation}\label{eq-cf}
\lambda_{\mathrm{cf}}=\lambda_{\mathrm{cf}}(\Sigma,D)=\lim_m \frac{m}{E_m[l]}\:.
\end{equation}
Let $p= (p(c))_{c\in {\mathfrak C}}$ be defined by $p=\lim_n \vec{1} (I+\widehat{A}+\cdots + \widehat{A}^{n-1})/n$. 
It can be interpreted as the limit distribution of the Markov chain $(X_n)_n$. 
According to the ergodic theorem for Markov chains (Theorem 4.6 in \cite{sene}), the limit
exists in \eref{eq-cf} and we have
\begin{equation}\label{eq-cf2}
\lambda_{\mathrm{cf}} \:  = \: (\sum_{c\in {\mathfrak C}} p(c) |c|)^{-1}\:.
\end{equation}

When  $(\Sigma,D)$ is connected, it follows from Lemma \ref{le-stco}
that $\widehat{A}$ is irreducible. Then $p$ is entirely determined by $p \widehat{A} = p$
and $\sum_i p_i=1$ (Perron-Frobenius Theorem, see \cite{sene}). 
It implies  that $\lambda_{\mathrm{cf}}$ is 
explicitly computable and rational.  
When $(\Sigma,D)$ is non-connected, 
$\lambda_{\mathrm{cf}}$ is still explicitly computable and rational according to Proposition \ref{le-critcf}.

\medskip

Consider an equitable partition  
$\pi=\{{\mathfrak C}_1,\dots ,{\mathfrak C}_s\}$ such that all the cliques 
in ${\mathfrak C}_i$ have a common length $l_i$.
There exists an analog of \eref{eq-cf2} corresponding to this partition. 
Let $\widehat{A}_{\pi}$ be the Markovian matrix associated with
the coloration matrix $A_{\pi}$. 
Let $p_{\pi}$ be defined by $p_{\pi}=\lim_n \vec{1} (I+\widehat{A}_{\pi}+\cdots + \widehat{A}_{\pi}^{n-1})/n$.
Then, we have 
$\lambda_{\mathrm{cf}}  =  (\sum_{i} p_{\pi}(i) l_i)^{-1}$.

\subsection{Non-connected dependence graphs}\label{sse-ncdg}

Assume that $(\Sigma,D)$ is {\em non-connected} and let 
$(\Sigma_s,D_s)_{s\in S}$ be the maximal connected subgraphs of $(\Sigma,D)$. 
We now propose formulas to express the average height of $(\Sigma,D)$
as a function of the ones of $(\Sigma_s,D_s)$.

\medskip

First, it is simple to prove using \eref{eq-king} and the Strong Law of Large Numbers (see also
Theorem 5.7 in \cite{sahe}) that we have 
\begin{equation}\label{eq-decomp}
\lambda_{*}(\Sigma,D) = \max_{s\in S} \ ( \ \frac{|\Sigma_s|}{|\Sigma|} \lambda_{*}(\Sigma_s,D_s) \ ) \:.
\end{equation}

\begin{proposition}\label{le-crit} 
Denote by $L_s$ the length generating function 
of $(\Sigma_s,D_s)$.
Define $J=\{j\in S, \ \rho_{L_j} = \min_{s\in S} \rho_{L_s}\}$. 
Then, we have 
\begin{equation}\label{eq-declambda}
\lambda_{\M}(\Sigma,D) = \lambda_{\M}(\Sigma_J,D_J)\:,
\end{equation}
where $\Sigma_J=\cup_{j\in J}\Sigma_j$, and $D_J=\cup_{j\in J}D_j$. 
\end{proposition}

The proof uses Proposition \ref{pr-lambda2} and is given in Appendix. 
There seems to be no simple way
to write $\lambda_{\M}(\Sigma_J,D_J)$ as a function of $\lambda_{\M}(\Sigma_j,D_j), j\in J$, as 
illustrated by the example of  section \ref{sse-fcm}. 

\begin{proposition}\label{le-critgamm}
Define $J=\{j\in S, \ |\Sigma_j|>1\}$. Then, we have 
\begin{equation}\label{eq-decgamma}
\gamma_{\M}(\Sigma,D) = \sum_{j\in J} \gamma_{\M}(\Sigma_j,D_j) + \frac{|S-J|}{2}\:,
\end{equation}
if $J\neq \emptyset$. If $J=\emptyset$, that is if $\M(\Sigma,D)$ is the free commutative monoid, 
we have $\gamma_{\M}(\Sigma,D) =(|\Sigma|+1)/2$.
\end{proposition}

The proof is given in Appendix. 
Proposition \ref{le-critgamm} is the counterpart of Proposition \ref{le-crit} for $\gamma_{\M}$, but it is
more precise.

\begin{proposition}\label{le-critcf}
Let $\widehat{A}$ be defined as in section \ref{sse-cf}. Let ${\mathfrak C}_s$ 
be the set of cliques of $(\Sigma_s,I_s)$. 
Define the matrix $B$ of dimension ${\mathfrak C}\times {\mathfrak C}$
as follows: $B_{ij}=\widehat{A}_{ij}$ if $i\not\in \bigcup_{s\in S} {\mathfrak C}_s$ and $B_{ij}=0$
otherwise. Define the vectors $\cI_{{\mathfrak C}_s},s\in S,$ of dimension ${\mathfrak C}$ as follows:
$(\cI_{{\mathfrak C}_s})_i=1$ if $i \in {\mathfrak C}_s$ and $(\cI_{{\mathfrak C}_s})_i=0$ otherwise. 
Set $q_s=\vec{1} (I-B)^{-1}\cI_{{\mathfrak C}_s}$, where $\vec{1}=(1/|{\mathfrak C}|,\dots , 1/|{\mathfrak C}|)$. 
Then we have
\begin{equation}\label{eq-dec2}
\lambda_{\mathrm{cf}}(\Sigma,D)^{-1}= \sum_{s\in S} q_s \ \lambda_{\mathrm{cf}}(\Sigma_s,D_s)^{-1}\:.
\end{equation}
\end{proposition}

\begin{proof}
%Assume first that $(\Sigma , D)$ is {\em strongly connected}. 
%According to Lemma \ref{le-stco}, the graph
%$\Gamma$ is  strongly connected and $A$ is therefore irreducible.
%% and {\em aperiodic}.
%Let $p = (p(c))_{c\in {\mathfrak C}}$ be the unique probability distribution on ${\mathfrak C}$ 
%such that
%$p \widehat{A} = p$ (Perron-Frobenius
%theorem). 
%We have then $\lambda_{\mathrm{cf}}= (\sum_{c\in {\mathfrak C}} p(c) |c|)^{-1}$. 
%\medskip
The graph of cliques $\Gamma$ of $(\Sigma,I)$ can be decomposed in its maximal strongly connected subgraphs (mscs). 
Replacing each mscs by one node,
we define the {\em condensed} graph of $\Gamma$. The {\em final} mscs are the mscs without any successor 
in the condensed graph. According to Lemma \ref{le-stco}, the final mscs are 
precisely the ones with sets of nodes ${\mathfrak C}_s$ where $s\in S$. 

Remark that the non-negative matrix $B$ is such that $\sum_j B_{ij} <1$
for every $i\in {\mathfrak C}$. In particular, it implies that 
$(I-B)$ is invertible. 
Define $q_s=\vec{1} (I-B)^{-1}\cI_{{\mathfrak C}_s}$ for every $s\in S$. 

%Consider a Markov chain $(X_n)_n$ on the state space ${\mathfrak C}$ 
%with transition matrix $\widehat{A}$ and with initial distribution 
%$\vec{1}$. 
The quantities $q_s$ can be interpreted 
in terms of the Markov chain $(X_n)_n$
defined in section \ref{sse-cf}:
we have $q_s=\lim_n P\{ X_n \in {\mathfrak C}_s\}$ 
(Theorem 4.4 in \cite{sene}). 
Let $A_s$ be the restriction of $A$
to the index set $({\mathfrak C}_s \times {\mathfrak C}_s)$ 
and let $\widehat{A}_s$ be the Markovian matrix associated with $A_s$.
Let 
$\widehat{p}_s$ be the unique probability distribution on ${\mathfrak C}_s$ such that 
$\widehat{p}_s\widehat{A}_s=\widehat{p}_s$ (Perron-Frobenius Theorem, Chapter 1 in \cite{sene}).
According to the ergodic theorem for Markov chains (Theorem 4.2 in \cite{sene}), we have 
$\lambda_{\mathrm{cf}}(\Sigma_s,D_s)= (\sum_{c\in {\mathfrak C}_s} p_s(c)|c|)^{-1}$. 

Define the vector $p_s$ of dimension ${\mathfrak C}$ by $p_s(c)=\widehat{p}_s(c)$ if $c\in 
{\mathfrak C}_s$ and $p_s(c)=0$ otherwise. 
The vector 
$p= \sum_{s\in S} q_s p_s$
is then the unique limit distribution of $(X_n)_n$. 
By the ergodic theorem for Markov chains (Theorem 4.6 in \cite{sene}), we obtain \eref{eq-dec2}. 
\end{proof}

%There exist analogs of the formula \eref{eq-cf} and \eref{eq-cf2} 
%with $\widehat{A}_{\pi}$ playing the role of $\widehat{A}$. 
%For instance, in the strongly connected case, if $p_{\pi}$ is
%the probability vector such that $p_{\pi} \widehat{A}_{\pi} = p_{\pi}$, 
%then 
%$\lambda_{\mathrm{cf}}  =  (\sum_{i} p_{\pi}({\mathfrak C}_i) l_i)^{-1}$.

\subsection{Comparison between the different average heights}

In terms of computability, the simplest quantity is $\lambda_{\mathrm{cf}}$ and the 
most complicated one is $\lambda_{*}$. This is reflected by the fact that 
$\lambda_{\mathrm{cf}}$ is rational, that $\lambda_{\M}$ and $\gamma_{\M}^{-1}$ are algebraic, and that 
$\lambda_{*}$ is in general not algebraic, see for instance \eref{eq-d7}. 

\medskip

Another point of view is to compare the families of probability measures $(P_n)_n,(Q_n)_n,(\tilde{Q}_n)_n$, and 
$(R_n)_n$ associated respectively with $\lambda_{*}, \lambda_{\M}, \gamma_{\M}^{-1},$ and 
$\lambda_{\mathrm{cf}}$. A family of probability measures $(\mu_n)_n$ defined on $(\M_n)_n$ (or $({}_n\M)_n$) 
is said to be {\em consistent} if we have $\mu_m\{t\}=\mu_n\{v\ : \ \exists u, \ v=tu\}$ for all $m<n$.
In this case, there exists a unique probability measure on infinite traces 
whose finite-dimensional marginals are the probabilities $(\mu_n)_n$. Consistency is a natural and 
desirable property. Clearly the families $(P_n)_n$ and $(R_n)_n$ are consistent. On the other hand, 
the families $(Q_n)_n$ and $(\tilde{Q}_n)_n$ are not.

It is also interesting to look at the asymptotics in $n$ of the empirical distribution of 
$\{h(t)/|t|, t\in \M_n\}$ or $\{|t|/h(t), t \in {}_n\M\}$. 
For $a\in \R$, let $\delta_a$ denote the probability measure concentrated in $a$. 
It follows from \eref{eq-king} that we have
\[
\sum_t P_n\{t\} \delta_{h(t)/|t|} \ \longrightarrow \ \delta_{\lambda_{*}}\:, 
\] 
with the arrow standing for `convergence in distribution'.  
%\[
%P_n \left\{ h(\psi(w))/|w| \in [ \lambda_{*}-\varepsilon, \lambda_{*}+\varepsilon]  \right\}
%\stackrel{n}{\longrightarrow} 1\:.
%\]
Similarly, it follows from the ergodic theorem for Markov chains that we have
\[ 
\sum_t R_n\{t\} \delta_{h(t)/|t|} \ 
\longrightarrow \ \sum_{s\in S} q_s \delta_{\lambda_{\mathrm{cf}}(\Sigma_s,D_s)}\: ,
\] 
%\[ 
%R_n \ \{ h(t)/|t| \in \cup_{s\in S} 
%[\lambda_{\mathrm{cf}}(\Sigma_s,D_s)-\varepsilon, \lambda_{\mathrm{cf}}(\Sigma_s,D_s)+\varepsilon]\ \} 
%\stackrel{n}{\longrightarrow} 1\:, 
%\] 
the notations being the ones of section \ref{sse-cf}. There are no such concentration
results for $(Q_n)_n$ and $(\tilde{Q}_n)_n$. To check this, consider the case of the free commutative
monoid over two letters. We obtain easily that
\[
\sum_t Q_n\{t\} \delta_{h(t)/|t|}  \longrightarrow  U, \; \;
\sum_t \tilde{Q}_n\{t\} \delta_{|t|/h(t)}  \longrightarrow  V\:,
\]
where $U$ is the uniform distribution over the interval $[1/2,1]$ and $V$ is the 
uniform distribution over the interval $[1,2]$. 

\medskip

Consider two dependence graphs $(\Sigma,D_1)$ and $(\Sigma,D_2)$ with $D_1\subset D_2$. The intuition
is that $\M(\Sigma,D_1)$ should be `more parallel' than $\M(\Sigma,D_2)$. 
In accordance with this intuition, it is elementary to prove that 
$\lambda_{*}(\Sigma,D_1)\leq \lambda_{*}(\Sigma,D_2)$. However, the corresponding 
inequalities do not hold for $\lambda_{\M}$ and $\lambda_{\mathrm{cf}}$. 
Consider for instance the trace monoids over three or four letters whose
average heights are given in section \ref{se-234}. This raises some interesting issues 
on how to interpret these quantities. 
On the other hand, we conjecture that 
the inequality $\gamma_{\M}(\Sigma,D_1)^{-1}\leq \gamma_{\M}(\Sigma,D_2)^{-1}$ is 
satisfied.

%To illustrate these points, consider $\Sigma=\{1,2,3\}$ and let $D_1=\{(i,i),i\in \Sigma\}$, $D_2=
%D_1\cup \{(1,2),(2,1)\}$, $D_3=D_2\cup\{(2,3),(3,2)\}$, and $D_4=\Sigma\times \Sigma$. The values 
%of the average heights are given in the table. 

%\begin{center}
%\begin{tabular}{l|c|c|c|c} 
%&&&& \\ 
% & $\lambda_{*}$ & $\lambda_{\M}$ &   $\gamma_{\M}^{-1}$ &  $\lambda_{\mathrm{cf}}$ \\ 
%&&&& \\ \hline
%&&&& \\
%$(\Sigma,D_1)$  &  1/3 & 11/18 & 1/2  & 1 \\
%$(\Sigma,D_2)$  &  2/3 & 1 & 2/3 & 1 \\
%$(\Sigma,D_3)$  &  $(10 + \sqrt{5})/15$ &  $(7+\sqrt{5})/10$&  \;\;\; 9/11 \;\;\; & \;\;\; 8/9 \;\;\; \\
%$(\Sigma,D_4)$  &  1 &  1 & 1 & 1 \\
%&&&& \\ 
%\end{tabular}
%\end{center}

\section{Some Examples}

\subsection{The free commutative monoid}\label{sse-fcm}

Consider the dependence graph $(\Sigma,D)$ with $D=\{(u,u),u\in \Sigma\}$. 
%Let $(\Sigma,I)$ be the complete graph over $\Sigma$, that is $I=\{(u,v), u,v \in \Sigma ,u\neq v\}$. 
%The strongly connected subgraphs of $(\Sigma,D)$ are the trivial graphs consisting 
%of one node and a self-loop. 
The corresponding trace monoid $\M(\Sigma,D)$ is the free commutative monoid over the alphabet $\Sigma$,
which is isomorphic to $\N^{\Sigma}$. Set now $k=|\Sigma|$. 

\medskip

A direct application of \eref{eq-decomp} yields $\lambda_{*}=1/k$. 
Consider now $\lambda_{\mathrm{cf}}$. The final maximal strongly connected subgraphs
of $\Gamma$ are precisely the cliques of length 1. 
In particular, they are of cardinality 1.  
Applying the results in section \ref{sse-cf}, we get $\lambda_{\mathrm{cf}}=1$. 
%Obviously,  
%the quantity  $\lambda_{\mathrm{cf}}$ is not well adapted in this case.  

\medskip

Let us compute  $\lambda_{\M}$ and $\gamma_{\M}$. 
%Note first that Propositions \ref{le-crit} and 
%\ref{le-critgamm} do not apply. 
Using the methodology of sections \ref{se-hlgf} and \ref{se-aah} is feasible,
but there are simpler methods. 
Consider $\gamma_{\M}$ first. By a counting argument, we get
\begin{equation}\label{eq-lm}
L_m=\sum_{t\in {}_m\M} y^{|t|}=(1+y+\cdots + y^m)^k - (1+y+\cdots + y^{m-1})^k\:.
\end{equation}
We obtain  $|{}_m\M|=L_m(1)=(m+1)^k-m^k$ and $\sum_{t\in {}_m\M} |t|=L_m'(1)=
k(m+1)^km/2 -km^k(m-1)/2$. We deduce that $\gamma_{\M}^{-1}=2/(k+1)$.

\medskip

Let us now compute the average height $\lambda_{\M}$. 
The length generating function is $L=(1-y)^{-k}$. 
%It follows that $b_n \sim n^{k-1}/ (k-1)\!$. 
Applying a result
from Carlitz \cite{carl62}, we have 
\begin{eqnarray}\label{eq-carlitz}
G=(\partial F/\partial x)(1,y)&=&\sum_{(n_1,\dots ,n_k)} \max (n_1,\dots , n_k) y^{n_1+\cdots +n_k} \nonumber \\
& = & \frac{1}{(1-y)^{k+1}} \sum_{i=1}^k {k \choose i} 
\frac{(-1)^{i-1}y^i}{1+y+\cdots + y^{i-1}}\:.
\end{eqnarray}
Using \eref{eq-complex0},
we obtain 
\begin{equation}\label{eq-carlitz2}
\lambda_{\M} \ = \ \frac{1}{k} \left( \sum_{i=1}^k {k \choose i} 
\frac{(-1)^{i-1}}{i} 
\right)\ = \
\frac{1}{k} \left( 1 + \frac{1}{2}+\cdots + \frac{1}{k} \right)\:.
\end{equation}
The last equality is a classical identity for harmonic summations (see Chapter 6.4 in \cite{GKPa}). 
%It can be proved recursively using Pascal's triangle identity. 
Asymptotically in $k$, we have $\lambda_{\M} \sim \log(k) /k$. This is to be compared with 
$\lambda_{*}=1/k$ and $\gamma_{\M}^{-1}\sim 2/k$. 

\medskip

Consider now the trace monoid $\M(\Sigma,D)$ obtained as the direct product of the free monoids 
$\Sigma_1^*,\dots , \Sigma_k^*$, with $|\Sigma_1|=\cdots = |\Sigma_k|=c$ and $c>1$. 
Equivalently, the dependence graph is $(\Sigma,D)$ with $\Sigma=\cup_{i=1}^k \Sigma_i$, $D=\cup_{i=1}^k D_i$, and
$D_i=\Sigma_i\times \Sigma_i$ for all $i$. 
Clearly, we still have $\lambda_*=1/k$ and $\lambda_{\mathrm{cf}}=1$. The formulas
in \eref{eq-lm} and \eref{eq-carlitz} still hold when replacing $y$ by $cy$. We deduce that $\lambda_{\M}$ 
is still given by \eref{eq-carlitz2}. On the other hand, we have 
$\gamma_{\M}^{-1}= 1/k$, a value which can also be obtained 
using Proposition \ref{le-critgamm}. 
Hence the value of $\gamma_{\M}^{-1}$ does not depend on the value of $c$, $c>1$, 
and is different from the value obtained for $c=1$.

\subsection{The ladder graph}\label{sse-hsg}

In view of Proposition \ref{th-A},
the simplest sets of cliques are those with the
property that the clique partition according to the length is
equitable, so that the dimension of the corresponding coloration
matrix reduces to the maximal size of a clique. 
This holds if
the full automorphism group of $(\Sigma,D)$ or $\Gamma$ acts transitively on the sets of
cliques of the same length. 
%If the dependence graph is a {\em rank 3} graph (see \cite[Chap. 2]{CVL} for a definition), 
%the above property holds. 
This is in particular the case when the dependence graph is the {\em triangular graph}, i.e. the line 
graph of the complete graph $K_n$, or the {\em square lattice graph}, i.e.
the line graph of the complete bipartite graph $K_{n,n}$.

%A good source of examples is the class of {\em rank 3 graphs} 
%(see \cite[Chap. 2]{CVL} for a definition). 
%The triangular graphs $T_n$ ($T_n$ is the line graph of the complete graph $K_n$) 
%are rank $3$ graphs. 
%A similar example is the {\em square lattice graph} 
%$L_{2}(n)$ with vertex set $\{ 1, \dots , n \} \times \{ 1, \dots , n \}$. 
%Two distinct vertices are adjacent if they agree in one
%coordinate. Hence $L_{2}(n)$ is isomorphic to the line graph of the complete
%bipartite graph $K_{n,n}$. Its automorphism group is the wreath
%product ${\mathfrak S}_{n}[{\mathfrak S}_{2}]$ of the symmetric groups of degrees $n$ and $2$ 
%and the corresponding coloration matrix is of
%dimension $n$. %Results analogous
%%to the ones of section \ref{se-tg} can be expected when $(\Sigma, D) = L_{2}(n)$.

\medskip

A particularly simple class of independence graphs is the class of 
node and arc-transitive triangle-free graphs. In this case, the coloration matrix 
associated with the full automorphism group is of dimension $2 \times 2$.
Let us consider a family of graphs of this type. 
%vertex and edge-transitive. 

\medskip

Let $(\Sigma, I)$ be the {\em ladder graph}, i.e. 
$\Sigma=\{1,\dots,2n\}$ and $I=\{(i,j)\in \Sigma\times \Sigma \ : \ i+j=2n+1\}$.
The corresponding dependence graph is known as the  
{\em cocktail party graph} $\mbox{CP}_n$. 
The full automorphism group is the wreath
product $W={\mathfrak S}_{n}[\Z/2\Z]$ of the symmetric group of degree $n$ with $\Z/2\Z$. 
The corresponding partition of ${\mathfrak C}$ is $\{{\mathfrak C}_1,{\mathfrak C}_2\}$ with 
${\mathfrak C}_1=\{c\in {\mathfrak C}, |c|=1\}$ and ${\mathfrak C}_2=\{c\in {\mathfrak C}, |c|=2\}$.
The coloration matrix is 
\[
A_W= \left(\begin{array}{cc} 2n-1 & n-1 \\ 2n & n \end{array}\right)\:.
\]
The computation of $\lambda_{*}(\mbox{CP}_n)$ was worked out by Brilman (see \cite{bril}, Proposition 14):
\[
\lambda_{*}(\mbox{CP}_n) =\frac{1}{2} \left( 1+ \frac{\sqrt{n-1}}{\sqrt{n+1}} \right)\:.
\]
We compute $F$ using the reduced representation induced by the 
partition. 
The dominant singularity of $L$ is $(1-\sqrt{1-n^{-1}})$ and we obtain
\[
\lambda_{\M}(\mbox{CP}_n)= \frac{1}{2}\left( 1 + \frac{\sqrt{n}}{2\sqrt{n} -\sqrt{n-1}} \right)\:.
\]
The dominant singularity of $H$ is 
$(3n-1-\sqrt{9n^2-10n+1})/2n$, and we get
\[
\gamma_{\M}(\mbox{CP}_n)^{-1}= 
\frac{\Delta(5n-1-\Delta)}{2\Delta(4n-1) - 2(8n^2-9n+1)}, \ \ \Delta= \sqrt{9n^2-10n+1}\:.
\]
Considering the Markovian matrix  $\widehat{A}_W$ and
using \eref{eq-cf2}, we obtain
\[
\lambda_{\mathrm{cf}}(\mbox{CP}_n)= \frac{9n-7}{12n-10}\:.
\]
We check that $\lim_{n} \lambda_{*}(\mbox{CP}_n)=\lim_{n} \lambda_{\M}(\mbox{CP}_n) =1$ and that
$\lim_{n} \gamma_{\M}(\mbox{CP}_n)^{-1}$ $=\lim_{n} \lambda_{\mathrm{cf}}(\mbox{CP}_n) = 3/4$.

\medskip

Remark that $\mbox{CP}_3 \equiv T_4$. By specializing the above results to $n=3$, we get the 
average heights for the triangular graph $T_4$ considered in 
Examples \ref{ex-k4}, \ref{ex-heapk4}, \ref{ex-cliqk4}, 
\ref{ex-colk4}, \ref{ex-colok4}, and \ref{ex-thk4}. We have
\begin{eqnarray*}
&\lambda_{*}(T_4)  \ = \ \frac{2+\sqrt{2}}{4} \ = \ 0.854\cdots\:, \ \  
\lambda_{\mathrm{cf}}(T_4) \ = \ \frac{10}{13} \ = \ 0.769\cdots \:, & \\
&\lambda_{\M}(T_4) \ = \ \frac{16 + \sqrt{6}}{20} \ = \ 0.922\cdots\:,\ \  
\gamma_{\M}(T_4)^{-1}  \ = \ \frac{39 + \sqrt{13}}{58} \ = \ 0.735\cdots\:. & 
\end{eqnarray*}

%\[
%\lambda_{*}(T_4)  =  \frac{1}{2}+ \frac{\sqrt{2}}{4}  = 0.85355\cdots\:, \ \ 
%\lambda_{\M}(T_4)  =  \frac{3\sqrt{6}-2}{4(\sqrt{6}-1)} =  0.92247\cdots\:,
%\]
%\[
%\gamma(T_4) =  \frac{10}{13}  =   0.76923\cdots \:.
%\]

%\begin{eqnarray*}
%\lambda_{*}(T_4) & = \  \frac{1}{2}+ \frac{\sqrt{2}}{4} & = \ 0.85355\cdots \\
%\lambda_{\M}(T_4) & = \ \frac{3\sqrt{6}-2}{4(\sqrt{6}-1)} & = \ 0.92247\cdots \\
%\gamma(T_4) & = \ \frac{10}{13} & =  \ 0.76923\cdots 
%\end{eqnarray*}

%For the triangular graphs $T_n, n\geq 5$
%(see section \ref{se-tg} for a definition), no simple close formula is known for $\lambda_{*}$.

%Assume now that $(\Sigma, I)$ is the {\em odd graph} $O_n$, with
%$n \geq 3$. Its vertices are the $(n-1)$-subsets of a fixed
%$(2n-1)$-set. Two vertices are adjacent if they are disjoint sets. 
%For $n=3$ we obtain the famous {\em Petersen graph} which is also the
%complement of $T_5$. Let $\overline{O_n}$ be the complement of
%$O_n$. Its full automorphism group is the symmetric group
%${\mathfrak S}_{2n-1}$. There is also an explicit, but a bit more complicated,
%formula for $\gamma(\overline{O_n})$ which yields 
%$\lim_{n \to \infty} \gamma(\overline{O_n}) = 1/2$.  

\appendix

\section{Proofs of the results in section \protect{\ref{sse-upt}}}

This section is devoted to the proof of Propositions \ref{pr-lambda2},
\ref{le-crit}, and \ref{le-critgamm}. 

\begin{proof}[Proof of Proposition \ref{pr-lambda2}] $ $ \\

We give the proof for $\lambda_{\M}$. The one for $\gamma_{\M}$ is similar (and easier!). 
Recall that 
$L=F(1,y)=\sum_{t\in \M} y^{|t|}$ is the length generating function. Define
\[
G=\frac{\partial F}{\partial x} (1,y) =\sum_{t\in \M} h(t) y^{|t|}\:.
\]
Assuming existence of the limit in \eref{eq-lambda2}, we have
$\lambda_{\M}=\lim_n (G|n)/(n(L|n))$. 

\medskip

According to Pringsheim's Theorem \cite[Sec. 7.21]{titc}, 
$L$ and $G$ have a positive real dominant singularity. They are  denoted respectively by
$\rho_L$ and $\rho_G$ according to the previous conventions. Since we have $n(L|n)/C \leq (G|n) \leq n(L|n)$,
it implies that $\rho_L=\rho_G$. 
Let $k_L$ be the order of $\rho_L$ in $L$.

\medskip

Assume that $\rho_L$ is the unique dominant singularity in $L$. 
Assume that the order of $\rho_L$ in $G$ is $(k_L+1)$ 
and is strictly larger than the one of the other
singularities of modulus $\rho_L$ (there might exist several dominant singularities for $G$, 
see section \ref{sse-fcm}). Then the limit in \eref{eq-lambda2} exists and we have, 
\[
\lambda_{\M} = \frac{ [G(y)(\rho_L-y)^{k_L+1}]_{|y=\rho_L}}{k_L\rho_L \cdot [L(y)(\rho_L-y)^k_L]_{|y=\rho_L}}\:.
\]
In particular, $\lambda_{\M}$ is an algebraic number. 
The above assumptions on the dominant singularities of $L$ and $G$ ensure that the sequences $((L|n))_n$ and 
$((G|n))_n$ do not have an oscillating behavior. 
It remains to prove that these assumptions 
actually hold. 

%Actually, as stated above, the series $L$ always have a unique 
%dominant singularity. On the other hand, as illustrated by the free commutative monoid in section 
%\ref{sse-fcm}, the series $G$ may have several dominant singularities. 
%Indeed, we are going to prove the following results. As The series $L$ has a unique dominant 
%singularity, its order is 1 when $(\Sigma,D)$ is strongly connected. 

\medskip

We work with the representation $(u,A(x,y),v(x,y))$ 
of $F$ given in the statement of Proposition \ref{th-A}. 
%Set $A=A(x,y)$ and $v=v(x,y)$. 
We have 
\begin{equation}\label{eq-f}
F=1+\frac{u\mbox{Adj}(I-A(x,y))v(x,y)}{\det(I-A(x,y))}\:,
\end{equation}
where $\det(.)$ stands for the determinant and Adj(.) for the adjoint of a matrix. 
Set $Q(x,y)= \det(I-A(x,y))$ and $F=P(x,y)/Q(x,y)$.
It follows that we have $L=P(1,y)/Q(1,y)$. By differentiating $F$, we get
\begin{equation}\label{eq-g}
G= \frac{1}{Q(1,y)} \left( \frac{\partial P}{\partial x}(1,y) - L\cdot \frac{\partial Q}{\partial x}(1,y) \right)\:.
\end{equation}
Set $Q(y)=Q(1,y)$. 
The above equations imply that the 
set of singularities of $L$ (resp. $G$) is included in the set of singularities of $1/Q(y)$. 
In particular, a dominant singularity of $L$ (resp. $G$) has a greater modulus than a dominant singularity 
of $1/Q(y)$.
%It implies that 
%$L=1+u(I-A(1,y))^{-1}v(1,y)$. 
%As usual we view the triple $(u,A(1,y),v(1,y))$ 
%as an automaton with multiplicities,
%i.e. as a weighted graph with input and output arcs.
%The coefficient $(L|n)$
%can be interpreted combinatorially as the number of paths with weight $y^n$ in the 
%automaton.
%Given an index set $S$ and $s\in S$, define $\cI_s\in \N^{1\times S}$ by 
%$(\cI_s)_s=1$ and $(\cI_s)_t=0, t\neq s$. 
%For $i,j\in {\mathfrak C}$, set 
%\[
%L_{ij}= \cI_i (I-A(1,y))^{-1}y^{|j|}\cI_j^T\:.
%\]
%The coefficient $(L_{ij}|n)$ can be interpreted as the number 
%of paths from $i$ to $j$ with weight $y^n$. 
%It follows, classically and straightforwardly, that $\sum_k L_{ik}L_{kj}$ can be interpreted as the sum
%of the {\it lengths} (number of arcs) of the paths from $i$ to $j$ with weight $y^n$.
%We deduce that
%\begin{equation}\label{eq-formula}
%L= 1+\sum_{ij} L_{ij}, \;\; G =\sum_{i,j} L_{ij} + \sum_{i,j,k} L_{ik}L_{kj}\:.
%\end{equation}
%The above formula can also be retrieved analytically. Using that $A=xA(1,y)$ and $v=xv(1,y)$,
%we obtain 
%\begin{eqnarray*}
%\frac{\partial F}{\partial x} & = & u (I-A)^{-1}\frac{\partial v}{\partial x} + 
%u \frac{\partial (I-A)^{-1}}{\partial x} v \\
% & = & u (I-A)^{-1} v(1,y) +  u (I-A)^{-1} \frac{\partial A}{\partial x} (I-A)^{-1} v \\
%&=& u (I-A)^{-1} v(1,y) +  u (I-A)^{-1} A(1,y) (I-A)^{-1} v \:.
%\end{eqnarray*}
%Hence, we get 
%\[
%G= u (I-A(1,y))^{-1} v(1,y) +  u (I-A(1,y))^{-1} A(1,y) (I-A(1,y))^{-1} v(1,y)\:,
%\]
%which is equivalent to the expression on the right in \eref{eq-formula}. 

\medskip

The next step consists in transforming the triple $(u,A(1,y),v(1,y))$ into another triple  
$(\tilde{u}, y\tilde{A}, y\tilde{v})$ of dimension $K>|{\mathfrak C}|$, 
where $\tilde{u}\in \N^{1\times K}, \tilde{A} \in \N^{K\times K}, \tilde{v}
\in \N^{K\times 1}$, and where we set $y\tilde{A}=(y\tilde{A}_{ij})_{ij}$ and 
$y\tilde{v}=(y\tilde{v}_i)_i$. 

\medskip

Before formally defining it, we illustrate the construction on the figure below. 
As usual we view a triple
as an automaton with multiplicities,
i.e. as a weighted graph with input and output arcs.
We have represented the portion of the automata  $(u,A(1,y),v(1,y))$ and 
$(\tilde{u}, y\tilde{A}, y\tilde{v})$ corresponding to
the cliques 
$u$ and $v$ where $|u|=3, |v|=2$, and $(u,v)$ is CF-admissible. 
\begin{figure}[htb]
\[ \epsfxsize=280pt \epsfbox{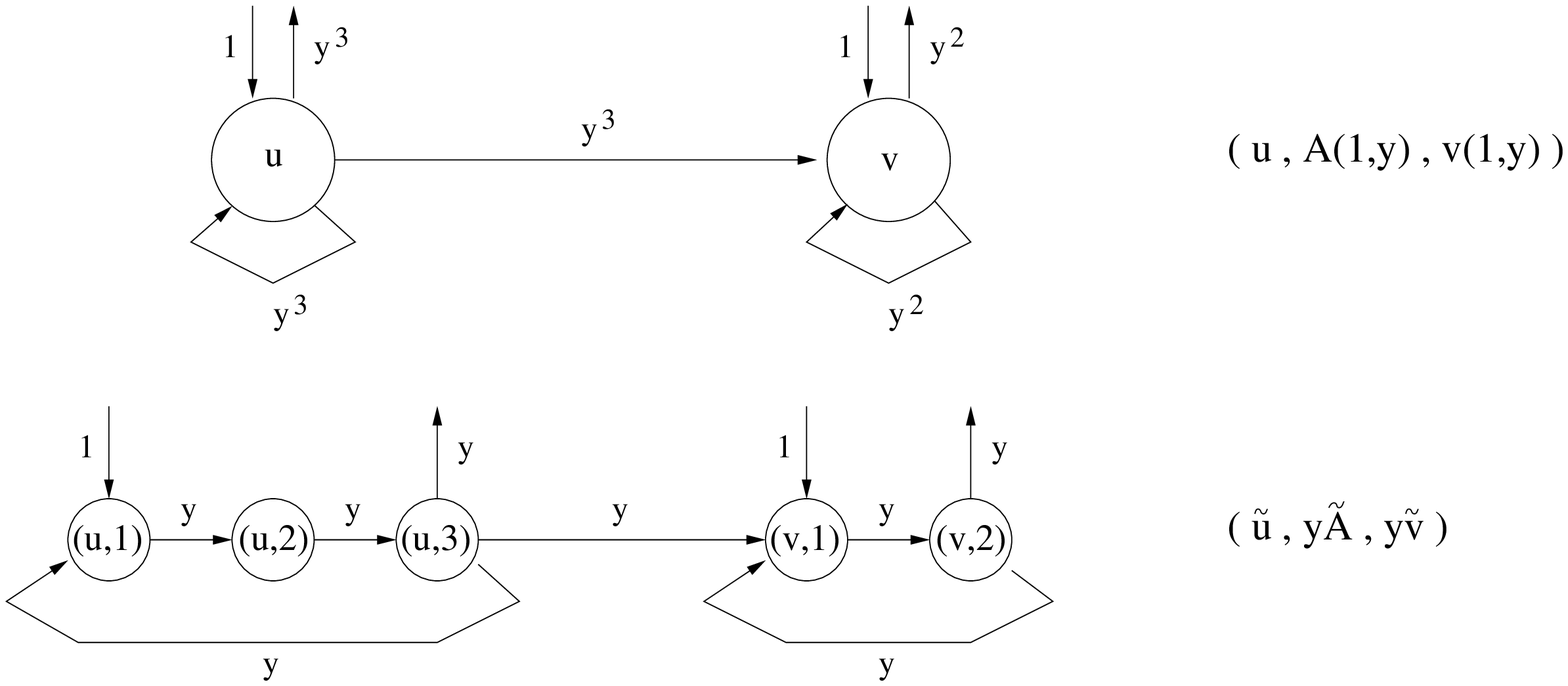} \]
\label{fi-transfo}
\end{figure}

Consider the index set 
\begin{equation}
E=\{(c,1),\dots , (c,|c|), c\in {\mathfrak C} \}\:.
\end{equation}
Let us define $\tilde{u}\in \N^{1\times E}, \tilde{A}\in \N^{E\times E},$ and $\tilde{v}\in \N^{E\times 1}$ 
as follows:
\begin{eqnarray*}
\tilde{u}_i & = & \begin{cases} 1 & \mbox{if } i=(c,1), c\in {\mathfrak C} \\
                            0 & \mbox{otherwise}\:, \end{cases} \\
\tilde{A}_{ij} & = & \begin{cases} 1 & \mbox{if } i=(c,k), j=(c,k+1), c\in {\mathfrak C}, 1\leq k < |c| \\ 
                                   1 & \mbox{if } i=(c,|c|), j=(c,1), c\in {\mathfrak C} \\
                                   1 & \mbox{if } i=(c,|c|), j=(d,1), A_{cd}\neq 0 \\
                                   0 & \mbox{otherwise}\:, \end{cases} \\
\tilde{v}_i & = & \begin{cases} 1 & \mbox{if } i=(c,|c|), c\in {\mathfrak C} \\
                            0 & \mbox{otherwise}\:. \end{cases} 
\end{eqnarray*}

In an automaton, an {\em input} (resp. {\em output}) {\em node} is a node with an input (resp. output) arc.
A {\em successful path} is a path from an input node to an output node. 
There is a one to one mapping between successful paths in the automata $(u,A(1,y),v(1,y))$ 
and $(\tilde{u}, y\tilde{A}, y\tilde{v})$:
to the successful path $(c_1,\dots , c_k)$ in $(u,A(1,y),v(1,y))$ corresponds the
successful path $((c_1,1),\dots , (c_1,|c_1|),(c_2,1),\dots ,$ $ (c_k , |c_k|))$ in 
$(\tilde{u},y\tilde{A},y\tilde{v})$, and vice versa. Note that the lengths of 
corresponding paths do not coincide. 
Using this correspondence, we get that
\begin{equation}\label{eq-L2}
L= 1+ u(I-A(1,y))^{-1}v(1,y) = 1+ \tilde{u}(I- y \tilde{A})^{-1} y\tilde{v}^T\:.
\end{equation}
Let us prove that 
\begin{equation}\label{eq-egaldet}
Q(y)=\det (I-A(1,y))= \det (I-y\tilde{A})\:.
\end{equation}
Given a matrix $M$ of dimension $n$, we have 
\[
\det(M) = \sum_{\sigma\in {\mathfrak S}_n} \mathrm{sgn}(\sigma) M_{1\sigma(1)}\cdots 
M_{n\sigma(n)}\:,
\]
where ${\mathfrak S}_n$ is the set of permutations of $\{1,\dots ,n\}$, and where $\mathrm{sgn}(.)$ is the sign 
of a permutation. The permutations having a non zero contribution to the determinant are the ones
which correspond to a partition into simple cycles of the nodes of the graph of $M$. 

We have seen above that there is a one-to-one correspondence between successful paths in the graphs of
$A(1,y)$ and $y\tilde{A}$. 
There is also clearly a one-to-one correspondence between simple cycles in (the graphs of)
$A(1,y)$ and $y\tilde{A}$. When comparing the simple cycles of $(I-A(1,y))$ and $(I-y\tilde{A})$, 
one needs to be more careful. 

Let $S$ be the set of simple cycles of $(I-A(1,y))$
and let $\tilde{S}$ be the one of $(I-y\tilde{A})$. 
To the simple cycle
$c=(c_1,\dots ,c_k)$ in $S$, there corresponds the simple 
cycle $\tilde{c}=((c_1,1),\dots, (c_1,|c_1|),(c_2,1),\dots , (c_k,|c_k|))$
in $\tilde{S}$.
A simple enumeration shows that 
\[
\tilde{S}=\{\tilde{c}, c\in S\} \cup \{((c,i)), c\in {\mathfrak C}, |c|>1, 1\leq i \leq |c|\}\:. 
\] 
Given $c\in S$ (resp. $\tilde{S}$), we denote by 
$w(c)$ the contribution of $c$ to $\det(I-A(1,y))$ (resp. $\det(I-y\tilde{A})$). More precisely,
for $c=(c_1,\dots , c_k)$
and setting $M=I-A(1,y)$ (resp. $M=I-y\tilde{A}$), we set
\[
w(c)= \sum_{\sigma\in {\mathfrak S}_k} \mathrm{sgn}(\sigma) M_{c_1c_{\sigma(1)}}\cdots M_{c_kc_{\sigma(k)}}\:.
\]
Consider $c=(c_1,\dots , c_k)\in S$. 
We have
\[
w(c)= \begin{cases}  1- y^{|c_1|} & \mbox{ if } k=1 \\
              (-1)^{k-1} \prod_{i=1}^k -y^{|c_i|} = - y^{\sum_i |c_i|} & \mbox{ if } k>1 
\end{cases} \:.
\]
Let $\tilde{c}=((c_1,1),\dots, (c_1,|c_1|),\dots , (c_k,|c_k|))$  
be the corresponding cycle of $\tilde{S}$. 
Then we have
\[
w(\tilde{c})= \begin{cases} 1- y & \mbox{ if } k=1 \mbox{ and } |c_1|=1 \\
(-1)^{\sum_i |c_i| -1} \prod_{i=1}^k \prod_{j=1}^{|c_i|} -y = - y^{\sum_i |c_i|} & \mbox{ otherwise } 
\end{cases} \:.
\]
We check that $w(c)=w(\tilde{c})$ except in the case $k=1, |c_1|>1$. In this last situation, we have
$w(c)= 1- y^{|c_1|}$ and $w(\tilde{c})= - y^{|c_1|}$. However,
this difference is precisely compensated by the contribution to
$\det(I-y\tilde{A})$ of the simple cycles in $\{((c,i)), c\in {\mathfrak C}, |c|>1, 1\leq i \leq |c|\}$. 
We conclude
that $\det (I-A(1,y))= \det (I-y\tilde{A})$. 

\medskip

Let $(\Sigma_i,D_i), i\in \cU,$ 
be the maximal connected subgraphs of $(\Sigma,D)$.
Let ${\mathfrak C}$ be the set of cliques of $(\Sigma,I)$ and let ${\mathfrak C}_i$ be the one
of $(\Sigma_i,I_i)$. 

For $V \subset \cU$, define ${\mathfrak C}_V=\{ \prod_{v\in V} c_v , c_v \in {\mathfrak C}_v\}$.
Note that we have ${\mathfrak C}_i= {\mathfrak C}_{\{i\}}$.
The set ${\mathfrak C}$ is partitioned by the sets ${\mathfrak C}_V, V \subset \cU$.
Let $\Gamma$ be the graph of cliques of $\M(\Sigma,D)$.
Using Lemma \ref{le-stco}, we get that the maximal strongly connected subgraphs of $\Gamma$ are 
the subgraphs with sets of nodes ${\mathfrak C}_V, V \subset \cU$. 
Clearly, there is a path in $\Gamma$ from a node in ${\mathfrak C}_V$ to a node in ${\mathfrak C}_W$ if and 
only if $W\subset V$. 

It implies the following. 
The restriction of the matrix $A$ to the index set ${\mathfrak C}_V$, denoted by $A_V$,
is irreducible. 
Now range the index set
${\mathfrak C}$ according to the order ${\mathfrak C}_{U_1},\dots , {\mathfrak C}_{U_k},$ where 
$U_1,\dots, U_k$, is an ordered list of the subsets of $\cU$ satisfying the property:
$U_i\subset U_j \implies i\geq j$. Then the matrix $A$ 
is block upper-triangular
with the blocks $A_{U_1},\dots ,A_{U_k},$  on the diagonal.
An analog statement holds for $\tilde{A}$, replacing ${\mathfrak C}_V$ by 
$\tilde{\mathfrak C}_V=\{(c,1),\dots , (c,|c|), c\in {\mathfrak C}_V \}$. We denote by $\tilde{A}_V$
the restriction of $\tilde{A}$ to the index set $\tilde{{\mathfrak C}}_V$. 
We have
\begin{eqnarray}\label{eq-det}
\det (I-A(1,y)) & = & \prod_{V\subset \cU} \det (I-A_V(1,y)) \nonumber\\ 
\det (I-y\tilde{A}) & = & \prod_{V\subset \cU} \det (I-y\tilde{A}_V)\:.
\end{eqnarray}

Given an index set $S$ and $s\in S$, define $\cI_s\in \N^{1\times S}$ by 
$(\cI_s)_s=1$ and $(\cI_s)_t=0, t\neq s$. 
For $i,j\in {\mathfrak C}$, define
\[
L_{ij}= \cI_i (I-A(1,y))^{-1}y^{|j|}\cI_j^T\:.
\]
The coefficient $(L_{ij}|n)$ can be interpreted combinatorially as the number 
of paths from $i$ to $j$ with weight $y^n$
in the automaton $(u,A(1,y),v(1,y))$.
In particular, we have $L=1+\sum_{i,j} L_{ij}$. 
With a proof similar to the one of \eref{eq-L2}, we get 
\begin{equation}\label{eq-L3}
L_{ij}= \cI_i(I-A(1,y))^{-1}y^{|j|}\cI_j^T = \cI_{(i,1)}(I- y \tilde{A})^{-1} y\cI_{(j,|j|)}^T\:.
\end{equation}

Consider $u \in {\mathfrak C}$. 
Let $A(1,y)_{[u]}$ denote the matrix obtained from $A(1,y)$ by replacing the line and the column $u$
by a line and a column of zeros. Then we have
\begin{equation}\label{eq-l1}
L_{uu}= \frac{y^{|u|}\mbox{Adj}(I-A(1,y))_{uu}}{\det(I-A(1,y))}=
\frac{y^{|u|}\det(I-A(1,y)_{[u]})}{\det(I-A(1,y))} \:.
\end{equation}
Let $\tilde{A}_{[u]}$ denote the matrix obtained from $\tilde{A}$ by replacing the line $(u,|u|)$ and the column 
$(u,1)$ by a line and a column of zeros. 
With a proof similar to the one of \eref{eq-egaldet}, we get 
\begin{equation}\label{eq-eg}
y^{|u|}\det(I-A(1,y)_{[u]})= y\det(I-y\tilde{A}_{[u]})\:.
\end{equation}
Assume that $u$ belongs to ${\mathfrak C}_U$ and let $\tilde{A}_{U,[u]}$
denote the restriction of $\tilde{A}_{[u]}$ to the index set $\tilde{{\mathfrak C}}_U$. 
Using \eref{eq-l1}, \eref{eq-egaldet}, \eref{eq-eg}, and \eref{eq-det}, we obtain
\begin{equation}\label{eq-numden}
L_{uu}= \frac{y\det(I-y\tilde{A}_{[u]})}{\det (I -y\tilde{A})}
 = \frac{y\det (I - y\tilde{A}_{U,[u]})}{\det (I -y\tilde{A}_U)} \:.
\end{equation}
We have $\tilde{A}_{U,[u]}\leq \tilde{A}_U$ 
(for the coordinate-wise ordering) and $\tilde{A}_{U,[u]}\neq \tilde{A}_U$. 
We have seen above that $\tilde{A}_U$ is irreducible. 
According to the Perron-Frobenius Theorem for irreducible matrices (see for instance \cite{sene}, Chapter 1.4),
it implies that the spectral radius of $\tilde{A}_{U,[u]}$ is strictly less than the one of $\tilde{A}_U$. 
Now, the roots of the polynomial $\det (I - y \tilde{A}_{U,[u]})$, resp. $\det (I -y\tilde{A}_U)$, 
are the inverses of the non-zero eigenvalues of $\tilde{A}_{U,[u]}$, resp. $\tilde{A}_{U}$.
Hence the possible simplifications between the numerator and the denominator in the right-hand side of 
\eref{eq-numden} do
not involve any dominant singularity.  

We conclude that the dominant singularities of $L_{uu}$ are precisely
the dominant singularities of $1/\det (I -y\tilde{A}_U)$. 
%Their 
%modulus is equal to the inverse of the spectral radius of $\tilde{A}_U$. 

\medskip

We have $(L|n) \geq (L_{uu}|n)$ for all $n$.  It implies that a dominant singularity 
of $L$ has a smaller modulus than a dominant singularity of $L_{uu}$. 
We deduce that a dominant singularity 
of $L$ has a smaller modulus than a dominant singularity of $1/\det (I -y\tilde{A}_U)$ for all $U$, hence 
a smaller modulus
than a dominant singularity of $1/Q(y)$. 
Using that $(G|n)\geq (L|n)$, 
we obtain the same result for $G$. 

\medskip

We conclude that the modulus of the dominant singularities of $L$, $G$, and $1/Q(y)$ 
are equal. Furthermore, the sets of dominant singularities of $L$ and $G$ are included in
the set of dominant singularities of $1/Q(y)$. 
Since $Q(y)=\det(I-y\tilde{A})$, the set of dominant singularities of $1/Q(y)$ is also equal to the set 
of inverses of maximal eigenvalues of $\tilde{A}$.  Let $\rho(\tilde{A})=\rho_L^{-1}$ denote the spectral radius 
of $\tilde{A}$. 

\medskip

First assume that $\rho(\tilde{A})=\rho_L=1$. 
According to Lemma \ref{le-rho=1}, $\M(\Sigma,D)$ is the free commutative monoid over $\Sigma$. 
The analysis
of section \ref{sse-fcm} applies. In particular, the limit $\lambda_{\M}$ in \eref{eq-lambda2}
exists and is given in \eref{eq-carlitz2}. It is obviously algebraic and even rational. 
Hence Proposition \ref{pr-lambda2} is satisfied in this case. 

\medskip

From now on, we assume that $\rho(\tilde{A})>1$. 
Let us specialize for a moment to the case where $(\Sigma,D)$ is connected. Using the above analysis,
the matrix $\tilde{A}$ is irreducible. For any $a\in \Sigma$, we have $\tilde{A}_{(a,1)(a,1)}>0$. We 
conclude that $\tilde{A}$ is primitive. By Perron-Frobenius 
Theorem for primitive matrices 
(\cite{sene}, Chapter 1.1), the matrix $\tilde{A}$ 
has a unique eigenvalue of maximal modulus which is positive real and of multiplicity 1. 

\medskip

We conclude that $\rho_L$ is the unique dominant singularity of $L$ and $G$. 
We conclude also that the order of $\rho_L$ is 1 in $L$, and at most 2 in $G$. 
Since $nC^{-1}(L|n)\leq (G|n) \leq n(L|n)$, we deduce that the order of
$\rho_L$ in $G$ is 2. 
 
We have just proved that the result of Proposition \ref{le-clique} holds for $L$. 
The proof of Proposition \ref{le-clique} for $H$ is similar (and easier). 

\medskip

Let us come back to the general case for $(\Sigma,D)$. 
Since we have now proved Proposition \ref{le-clique}, we are allowed to use Proposition \ref{le-clique2}
(the proof of the latter requires the former). 
We conclude that in all cases, $L$ has a unique dominant singularity.

\medskip

It remains to study the set of dominant singularities of $G$. To do this,
we study the set of eigenvalues of $\tilde{A}$ of maximal modulus.

\medskip

Fix  a subset $V \subset \cU$ and consider the restricted matrix $\tilde{A}_V$. 
Let $\rho(\tilde{A}_V)$ denote the spectral radius of $\tilde{A}_V$. We distinguish between 
two cases. 

\medskip

{\em Case (I)}.
Assume there exists
$v\in V$ such that $\M(\Sigma_v,D_v)$ is different from the free monoid $\Sigma_v^*$, or equivalently 
such that $I_v$ is not empty. Then there exists $c,d\in {\mathfrak C}_V$ such that $|c|=|d|+1$. 
It implies that the cyclicity of the matrix $\tilde{A}_V$ is 1. Since $\tilde{A}_V$ is irreducible, we deduce
that it is primitive. According to the Perron-Frobenius 
Theorem for primitive matrices 
(\cite{sene}, Chapter 1.1), the matrix $\tilde{A}_V$ 
has a unique eigenvalue of maximal modulus which is positive real and of multiplicity 1. 
%Since $\tilde{A}_V$ is a $\{0,1\}$-matrix, we have 
%\begin{equation}\label{eq-cas1}
%\rho(\tilde{A}_V) >1\:.
%\end{equation}

\medskip

{\em Case (II)}.
Assume now that $\M(\Sigma_v,D_v)=\Sigma_v^*$ for all $v\in V$. It implies that 
$|c|=|V|$ for all $c\in {\mathfrak C}_V$. The cyclicity of $\tilde{A}_V$ is $|V|$ and 
$\tilde{A}_V$ is not primitive as soon as $|V|>1$. However in this case, we are able to
completely compute the spectrum of $\tilde{A}_V$. Set $K=\prod_{v\in V} |\Sigma_v|$.
It is more convenient to work with $A_V(1,y)$. 
Using the same arguments as in the proof of \eref{eq-egaldet},
we get
\[
\det(I-A_V(1,y))= \det(I-y\tilde{A}_V) \:.
\]
We also have $A_V(1,y) = y^{|V|}A_V(1,1)$ and $A_V(1,1)$ is the matrix of 
dimension $K\times K$ whose entries are all equal to 1. The eigenvalues of $A_V(1,1)$ are 0 with 
multiplicity $(K-1)$ and $K$ with multiplicity 1. We have
\begin{eqnarray*}
\det(I-y^{|V|}A_V(1,1)) & = & y^{K|V|}\det(y^{-|V|}I-A_V(1,1)) \\
& = & y^{K|V|}(y^{-|V|})^{K-1}(y^{-|V|}-K)=(1-Ky^{|V|})\:.
\end{eqnarray*}
It follows that the non-zero eigenvalues of $\tilde{A}_V$ are 
\begin{equation}\label{eq-cas2}
K^{1/|V|}\exp \ (\frac{2i\pi k}{|V|}), \ k=0,\dots, |V|-1,
\end{equation}
all with multiplicity 1. In particular, we have $\rho(\tilde{A}_V)=K^{1/|V|}$.
According to \eref{eq-det}, the spectral radius of $\tilde{A}$ 
is given by
\[
\rho(\tilde{A}) =\max_{V\subset \cU} \rho(\tilde{A}_V)\:.
\]
Define $\cV =\{ V\subset  \cU : \forall v\in V, I_v=\emptyset\}$ and 
$\cS=\{V\subset \cU : \rho(\tilde{A}_V)=\rho(\tilde{A})\}$.  
Using the above analysis, we can distinguish between two situations.

\medskip

First, assume that $\cV \cap \cS=\emptyset$.
According to Case (I), 
it implies that $\rho(\tilde{A})$ is the only eigenvalue of maximal modulus of the matrix $\tilde{A}$.
We conclude  that $\rho_G=\rho_L=\rho(\tilde{A})^{-1}$ is the only dominant singularity of $G$. 

Second, assume that there exists $U\in \cV \cap \cS$.
According to Case (II), it implies that 
\begin{equation}\label{eq-interm}
\rho(\tilde{A}) = \rho(\tilde{A}_U)=( \ \prod_{u\in U} {|\Sigma_u|} \ )^{1/|U|}\:.
\end{equation}
%Since $\rho(\tilde{A})>1$, it follows that 
%there exists $v\in V$ such that $|\Sigma_v|>1$. Since $\rho(\tilde{A}_V) \geq \rho(\tilde{A}_W)$ 
%for all $W\in \cV$, 
Let $\cP(S)$ denote the power set of a set $S$. 
We deduce easily from \eref{eq-interm} that 
\[
\cV \cap \cS = \cP(V), \ \mathrm{ with } \ V = \{v\in \cV \ : \  |\Sigma_v|=\max_{x\in \cV} |\Sigma_x|\}\:.
\]
We conclude that the set of maximal eigenvalues of $\tilde{A}$ is precisely 
given by 
\begin{equation}\label{eq-eigen}
\bigcup_{L=1}^{|V|} \left\{(\max_{x\in \cV} |\Sigma_x|) \exp \ (2i\pi l/L), \ l=0,\dots, L-1\right\}\:.
\end{equation}
Set $K=\max_{x\in \cV} |\Sigma_x|$. 
The multiplicity of the eigenvalue $K$ is at least $|V|$ (and it is exactly $|V|$ if
$\cS=\cP(V)$). 
For a complex and non positive real maximal eigenvalue, 
the multiplicity is exactly the number of appearances 
of the eigenvalue in \eref{eq-eigen}.
The maximal such multiplicity 
is equal to $\lfloor |V|/2 \rfloor$ and attained for the
eigenvalue $-K$. 
It follows that the maximal order of a complex and non positive real dominant singularity in
$1/Q(y)$ is $\lfloor |V|/2 \rfloor$. 
Using \eref{eq-g}, we conclude that the maximal order of a complex and 
non positive real dominant singularity in
$G$ is $\lfloor |V|/2 \rfloor$. 

\medskip

Now we also have 
$L=\prod_{v\in V} L_v \cdot \prod_{v \not\in V} L_v$. 
The M\"obius function of $(\Sigma_v,I_v), v\in V,$ is $(1-Ky)$. We deduce that
\[
L= \frac{1}{(1-Ky)^{|V|}}\cdot \prod_{v \not\in V} L_v\:.
\]
The order of the singularity $1/K$ in $L$ is consequently at least $|V|$. 
Since we have $n(L|n)/C \leq (G|n) \leq n(L|n)$, we deduce 
that one of the dominant singularities of $G$ must be of order $(k_L+1)\geq |V|+1$. 
Since we have $\lfloor |V|/2 \rfloor < |V|+1$, the only possible choice is $1/K$. 

\medskip

We conclude that the positive real dominant singularity of $G$ has a strictly larger order 
than all the other dominant singularities. It completes the proof.
\end{proof}

\begin{proof}[Proof of Proposition \protect{\ref{le-crit}}] $ $ \\

The notations are borrowed from the statement of  Proposition \ref{le-crit}.
To avoid trivialities, assume that $J\neq S$. 
Let $(\Sigma_1,D_1)=\cup_{j\in J} (\Sigma_j,D_j),$ and
$(\Sigma_2,D_2)=\cup_{j\in (S - J)} (\Sigma_j,D_j)$. 
Let $L$, $L_1$, and $L_2$ be the respective length generating functions of $(\Sigma,D)$, 
$(\Sigma_1,D_1)$, and  $(\Sigma_2,D_2)$.
%Let $\rho_1$ and $\rho_{L_2}$ be the respective dominant singularities
%of $L_1$ and $L_2$.
By construction, we have $\rho_{L_1}<\rho_{L_2}$. Let $k_{L_1}$ and $k_{L_2}$ be the order of $\rho_{L_1}$ and 
$\rho_{L_2}$ 
in their respective series. 

According to Proposition \ref{le-clique2}, we have
$(L_1|n)= a_n n^{k_1-1} \rho_{L_1}^{-n}$ with $\lim_n a_n$ $ =a \in \R_+^*$, and 
$(L_2|n)= b_n n^{k_2-1} \rho_{L_2}^{-n}$ with $\lim_n b_n=b \in \R_+^*$. 
Furthermore, we have $L=L_1L_2$, hence 
\begin{eqnarray*}
(L|n) & = & \sum_{i=0}^n a_{n-i} (n-i)^{k_1-1} \rho_{L_1}^{-(n-i)} b_i i^{k_2-1} \rho_{L_2}^{-i}  \\
  & = & a n^{k_1-1}\rho_{L_1}^{-n} 
\left(\sum_{i=0}^n (a_{n-i}/a) b_i i^{k_2-1}(1-i/n)^{k_1-1}(\rho_{L_1}/\rho_{L_2})^{i}\right) \:. 
\end{eqnarray*}
Since $\rho_{L_1}/\rho_{L_2} <1$, the series $B=\sum_{i=0}^{+\infty} b_i i^{k_2-1} (\rho_{L_1}/\rho_{L_2})^{i}$ is convergent.
Furthermore, we obtain easily that
\[
\lim_n \sum_{i=0}^n (a_{n-i}/a)b_i i^{k_2-1}(1-i/n)^{k_1-1}(\rho_{L_1}/\rho_{L_2})^{i} =B \:.
\]
We deduce that 
$(L|n)\sim aB n^{k_1-1} \rho_{L_1}^{-n}$.

\medskip

Let $f$ be an increasing  map from $\N$ to $\N$ such that $\lim_n f(n)=+\infty$ and $\lim_n f(n)/n=0$. 
Define 
\[
{\cal L}_n=\{ t \in \M(\Sigma,D), |t|=n, n -f(n)\leq |t|_{\Sigma_1}\leq n\}\:,
\]
where $|t|_{\Sigma_1}=\sum_{a\in \Sigma_1} |t|_a$. We have
\[
\# {\cal L}_n = a n^{k_1-1}\rho_{L_1}^{-n} 
\left(\sum_{i=0}^{f(n)} (a_{n-i}/a)b_i i^{k_2-1}(1-i/n)^{k_1-1}(\rho_{L_1}/\rho_{L_2})^{i}\right)\:.
\]
Since $\lim_n f(n)=+\infty$, we have $\# {\cal L}_n \sim aB n^{k_1-1} \rho_{L_1}^{-n}$ and 
$\lim_n \# {\cal L}_n/(L|n)=1$. 
Let $\M_n=\{t\in \M(\Sigma,D), |t|=n\}$ and note that $\# \M_n =(L|n)$. We have
\begin{eqnarray*}
\lambda_{\M}(\Sigma,D) & = & \lim_n \frac{\sum_{t \in {\cal L}_n}   h(t)
+ \sum_{t \in (\M_n - {\cal L}_n)}  h(t)}{n\cdot \# \M_n} \\ 
                       & = & \lim_n 
                             \underbrace{\frac{\sum_{t \in {\cal L}_n} h(t)}{n\cdot \# {\cal L}_n}}_{g_n} + 
\lim_n \frac{\sum_{t \in (\M_n - {\cal L}_n)} h(t)}{n\cdot \# \M_n} \:.
\end{eqnarray*}
Using the inequality $h(t)\leq |t|$, we obtain
\[
\frac{\sum_{t \in (\M_n - {\cal L}_n)} h(t)}{n\cdot \# \M_n}
 \leq   \frac{n\cdot( \# \M_n - \#  {\cal L}_n)}{n\cdot \# \M_n} \; \stackrel{n}{\longrightarrow} \; 0\:.
\]
We now consider the terms $g_n$. 
Given a trace $t$, we can decompose it as $t=\phi_1(t)\phi_2(t)$ with $\phi_1(t)\in \M(\Sigma_1,D_1)$ 
and $\phi_2(t) \in \M(\Sigma_2,D_2)$. 
Consider a trace $t$ such that $|t|=n$ and $|t|_{\Sigma_1} \geq n-f(n)$. 
We have $h(\phi_1(t))\geq C^{-1}(n-f(n))$ and $h(\phi_2(t)) \leq f(n)$, where $C$ is the maximal length of a 
clique. Using that $\lim_n f(n)/n=0$, we obtain that, for $n$ large enough, $h(t)=h(\phi_1(t))$. Hence,
we have, for $n$ large enough,
\begin{equation}\label{eq-intermed}
g_n  = 
\sum_{i=n-f(n)}^n \frac{\# \{|t|=n, |t|_{\Sigma_1} =i\}}{\# {\cal L}_n} \cdot
\frac{\sum_{|t|=n, |t|_{\Sigma_1} =i} \ h(\phi_1(t)) }{n\cdot \# \{|t|=n, |t|_{\Sigma_1} =i\}}\:.
\end{equation}
Given $u\in \M(\Sigma_1,D_1)$ and $n\geq |u|$, we have
\[
\# \{ t \in \M(\Sigma,D), |t|=n, \phi_1(t)=u \}= (L_2|n-|u|)\:,
\]
which depends on $u$ only via its length. 
We deduce that 
\begin{eqnarray*}
\frac{\sum_{t\in \M(\Sigma,D), |t|=n, |t|_{\Sigma_1} =i} \ h(\phi_1(t)) }
{n\cdot \# \{t\in \M(\Sigma,D), |t|=n, |t|_{\Sigma_1} =i\}}  & = &
\frac{i}{n}\cdot \frac{\sum_{t\in \M(\Sigma_1,D_1), |t| =i} \ h(t) }
{i\cdot \# \{t\in \M(\Sigma_1,D_1), |t|=i\}}\:. \\
& \sim & (i/n) \lambda_{\M}(\Sigma_1,D_1) \:.
\end{eqnarray*}
%By definition, the right hand side of the above equation converges as $n$
%goes to infinity to $\lambda_{\M}(\Sigma_1,D_1)$.
Replacing in \eref{eq-intermed}, we conclude that $\lambda_{\M}(\Sigma,D)=\lim_n g_n=\lambda_{\M}(\Sigma_1,D_1)$. 
\end{proof}

\medskip

\begin{proof}[Proof of Proposition \ref{le-critgamm}] $ $ \\

The notations are the ones of the statement of Proposition \ref{le-critgamm}. 
Assume first that $\M(\Sigma,D)$ is the free commutative monoid over $\Sigma$. 
According to the results of 
section \ref{sse-fcm}, 
we have indeed $\gamma_{\M}(\Sigma,D)=(|\Sigma| +1)/2$.

\medskip

Assume now that $\M(\Sigma,D)$ is not the free commutative monoid. Let $(\Sigma_2,D_2)$ be a maximal 
connected subgraph
of $(\Sigma,D)$ and let $\Sigma_1=\Sigma - \Sigma_2$ and $D_1=D-D_2$. 
Denote respectively by $H, H_1,$ and $H_2$ the height 
generating functions of $\M(\Sigma,D), \M(\Sigma_1,D_1)$, and $\M(\Sigma_2,D_2)$. 
We choose $(\Sigma_2,D_2)$ so that   $\M(\Sigma_1,D_1)$ is different
from the free commutative monoid. According to Lemma \ref{le-rho=1}, it implies that 
$\rho_{H_1}<1$. 
We are going to
prove the following equalities
\begin{equation}\label{eq-toprove}
\gamma_{\M}(\Sigma,D)=\begin{cases} \gamma_{\M}(\Sigma_1,D_1) + \gamma_{\M}(\Sigma_2,D_2) & 
\mbox{ if } |\Sigma_2|>1 \\ \gamma_{\M}(\Sigma_1,D_1) + 1/2 & \mbox{ if } |\Sigma_2|= 1 
\end{cases}\:.
\end{equation}
Formula \eref{eq-decgamma} follows easily from the above. 

\medskip

Assume first that $|\Sigma_2|>1$. According to Lemma \ref{le-rho=1}, it implies that 
$\rho_{H_2}<1$. 
Applying Propositions \ref{le-clique} and \ref{le-clique2}, 
we have $(H_1|n)=a_n n^{k_{H_1}-1} \rho_{H_1}^{-n}$ and 
$(H_2|n)=b_n \rho_{H_2}^{-n}$ with $\lim_n a_n=a$ and $\lim_n b_n=b$. 
Using \eref{eq-hh1h2} and performing the same type of computations as in the proof of Proposition 
\ref{le-clique2}, we get
\begin{equation}\label{eq-simh}
(H|n) \sim ab(\frac{1}{1-\rho_{H_1}} + \frac{1}{1-\rho_{H_2}} -1 ) n^{k_{H_1}-1}(\rho_{H_1}\rho_{H_2})^{-n}\:.
\end{equation}
We define the maps $f$, $\phi_1(.)$ and $\phi_2(.)$ as in the proof of Proposition \ref{le-crit}. 
Consider the set
\[
{\cal H}_n = \{ t\in \M(\Sigma,D), \ h(t)=n, \ h(\phi_1(t))\geq n-f(n), \ h(\phi_2(t))\geq n-f(n) \}\:.
\]
Using the same type of arguments as in the proof of Proposition \ref{le-crit}, it is easily seen that
$\lim_n \# {\cal H}_n/(H|n)=1$. Set ${}_n \M=\{t \in \M(\Sigma,D), h(t)=n\}$
and note that $\# {}_n \M =(H|n)$. 
We have
\begin{eqnarray*}
\gamma_{\M}(\Sigma,D) & = & \lim_n \frac{\sum_{t \in {\cal H}_n} |t|
+ \sum_{t \in ({}_n \M - {\cal H}_n)} |t|}{n\cdot \# {}_n \M } \\  
& = & \lim_n  \frac{\sum_{t \in {\cal H}_n} |t|}{n\cdot \# {\cal H}_n} + 
\frac{\sum_{t \in ({}_n \M - {\cal H}_n)} |t| }{n\cdot \# {}_n \M}\:.
\end{eqnarray*}
Using the inequality $|t|\leq C h(t)$, where $C$ is the maximal length of a clique, we obtain
\[
\frac{\sum_{t \in ({}_n \M - {\cal H}_n)} |t|}{n\cdot \# {}_n \M} \leq \frac{ C n (\# {}_n \M - \#{\cal H}_n)}{n\cdot 
\# {}_n \M} 
\stackrel{n}{\longrightarrow} 0\:.
\]
Using the equality $|t|=|\phi_1(t)|+ |\phi_2(t)|$, we obtain
\begin{eqnarray*}
\gamma_{\M}(\Sigma,D) & = & \lim_n \frac{\sum_{t \in {\cal H}_n} |\phi_1(t)|}{n\cdot \# {\cal H}_n}
 + \frac{\sum_{t \in {\cal H}_n} |\phi_2(t)|}{n\cdot \# {\cal H}_n} \\
& = & \gamma_{\M}(\Sigma_1,D_1) + \gamma_{\M}(\Sigma_2,D_2)\:,
\end{eqnarray*}
where the last equality is obtained exactly in the same way as in the proof of Proposition \ref{le-crit}. 

\medskip

Assume now that $|\Sigma_2|=1$. 
Then we have $\rho_{H_1}<1$ and $\rho_{H_2}=1$. 
It implies that $(H_1|n)\sim an^{k_{H_1}-1}\rho_{H_1}^{-n}$ and $(H_2|n)=1$.
Using \eref{eq-hh1h2}, we obtain that $(H|n)\sim n(H_1|n)$. 
Now, by a direct computation, we get, for $u\in \M(\Sigma_1,D_1)$,
\[
\#\{t \in \M(\Sigma,D), \phi_1(t)=u, h(t)=h(u)\}=h(u)+1\:.
\]
Define the set  ${\cal H}_n=\{t \in \M(\Sigma,D), h(t)=h(\phi_1(t))=n\}$. 
We have $\# {\cal H}_n = (n+1)(H_1|n) \sim (H|n)$. 
It implies that 
\begin{eqnarray*}
\gamma_{\M}(\Sigma,D) & = & \lim_n \frac{\sum_{t \in {\cal H}_n} |\phi_1(t)|}{n \cdot \# {\cal H}_n}
 + \frac{\sum_{t \in {\cal H}_n} |\phi_2(t)|}{n \cdot \# {\cal H}_n} \\
 &= & \lim_n \frac{(n+1)\sum_{t \in \M(\Sigma_1,D_1), h(t)=n} |t|}{n \cdot n(H_1|n)} +
\frac{\sum_{t \in \M(\Sigma_1,D_1), h(t)=n} \sum_{i=0}^n i }{n \cdot n(H_1|n) } \\
& = & \gamma_{\M}(\Sigma_1,D_1) + 1/2 \:.
\end{eqnarray*}
This completes the proof. 
\end{proof}

\newpage

\section{Trace monoids over 2, 3, and 4 letters}\label{se-234}

We give the values of the average heights for all the trace monoids over alphabets 
of cardinality 2, 3, and 4. On the tables below, a trace monoid is represented by its (non-directed) dependence graph. 
For readability, self-loops have been omitted in the dependence graphs. 
We have not represented the free monoids for which 
$\lambda_*=\lambda_{\M}=\gamma_{\M}=\lambda_{\mathrm{cf}}=1$.

\begin{center}
\begin{tabular}{|l|c|c|c|c|} \hline
&&&& \\ 
 & $\lambda_{*}$ & $\lambda_{\M}$ &   $\gamma_{\M}^{-1}$ &  $\lambda_{\mathrm{cf}}$ \\ 
&&&& \\ \hline\hline
\includegraphics{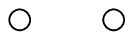}  &  \;\; \raisebox{2mm}{1/2} \;\;& \;\;\raisebox{2mm}{3/4} \;\;& \;\;\raisebox{2mm}{2/3}  \;\;& \;\;\raisebox{2mm}{1} \;\;\\ \hline
\end{tabular}

\bigskip

I. Trace monoids over 2 letters
\end{center}

The values in Table I can be obtained using the 
results in section \ref{sse-fcm}. 

\medskip

\begin{center}
\begin{tabular}{|l|c|c|c|c|} \hline
&&&& \\ 
 & $\lambda_{*}$ & $\lambda_{\M}$ &   $\gamma_{\M}^{-1}$ &  $\lambda_{\mathrm{cf}}$ \\ 
&&&& \\ \hline\hline
\includegraphics{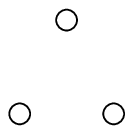}  &  \raisebox{6mm}{1/3} & \raisebox{6mm}{11/18} & \raisebox{6mm}{1/2}  & \raisebox{6mm}{1} \\ \hline
\includegraphics{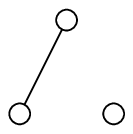}  &  \raisebox{6mm}{2/3} & \raisebox{6mm}{1} & \raisebox{6mm}{2/3} & \raisebox{6mm}{1} \\ \hline
\includegraphics{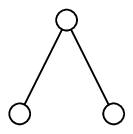}  &  \raisebox{6mm}{$(10 + \sqrt{5})/15$} &  \raisebox{6mm}{$(7+\sqrt{5})/10$} &  \;\;\; \raisebox{6mm}{9/11} \;\;\; & \;\;\; \raisebox{6mm}{8/9} \;\;\; \\ \hline
\end{tabular}

\bigskip

II. Trace monoids over 3 letters
\end{center}

All the values in Table II except one 
can be obtained using the results from the paper.
The exception is
$\lambda_{*}$ for $\Sigma=\{a,b,c\}, I=\{(b,c),(c,b)\}$,
which is computed in \cite{sahe}, Example 6.2. 

\medskip

\begin{center}
\begin{tabular}{|c|c|c|c|c|c|} \hline
&&&&& \\ 
& & $\lambda_{*}$ & $\lambda_{\M}$ &   $\gamma_{\M}^{-1}$ &  $\lambda_{\mathrm{cf}}$ \\
&&&&& \\ \hline\hline
\raisebox{6mm}1 & \includegraphics{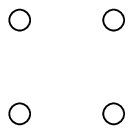} &  \raisebox{6mm}{$1/4$} &  \raisebox{6mm}{$25/48$}   
&  \raisebox{6mm}{$2/5$}  & \raisebox{6mm}{1}  
\\ \hline
\raisebox{6mm}2 & \includegraphics{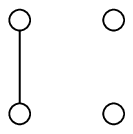} &  \raisebox{6mm}{$1/2$} &  \raisebox{6mm}{1}  
&  \raisebox{6mm}{$1/2$} & \raisebox{6mm}{1}  \\ \hline
\raisebox{6mm}3 & \includegraphics{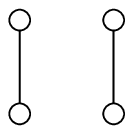} &  \raisebox{6mm}{$1/2$} &  \raisebox{6mm}{$3/4$}  
&  \raisebox{6mm}{$1/2$} & \raisebox{6mm}{1}  \\ \hline
\raisebox{6mm}4 & \includegraphics{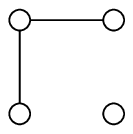} &  \raisebox{6mm}{$(10 + \sqrt{5})/20$} 
&  \raisebox{6mm}{$(7 + \sqrt{5})/10$} &  \raisebox{6mm}{$18/31$} &  
\;\; \raisebox{6mm}{$52/57$} \;\; \\ \hline
\raisebox{6mm}5 & \includegraphics{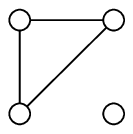} &  \raisebox{6mm}{$3/4$} &  \raisebox{6mm}{1}
&  \raisebox{6mm}{$2/3$} & \raisebox{6mm}{1}  \\ \hline
\raisebox{6mm}6 & \includegraphics{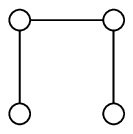} &  \raisebox{6mm}{?} & \raisebox{6mm}{$19/22$} 
&  \raisebox{6mm}{$(13 - 2\sqrt{13})/9$} &  \raisebox{6mm}{$5/6$} \\ \hline
\raisebox{6mm}7 & \includegraphics{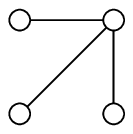} &  \raisebox{6mm}{in \eref{eq-d7}} & \raisebox{6mm}{in \eref{eq-l7}} 
&  \raisebox{6mm}{in \eref{eq-g7}} &  \raisebox{6mm}{$11/14$} \\ \hline
\raisebox{6mm}8 & \includegraphics{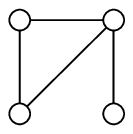} &  \raisebox{6mm}{$(5+\sqrt{2})/8$} 
&  \raisebox{6mm}{$(6+\sqrt{2})/8$} & \raisebox{6mm}{in \eref{eq-alg}} & \raisebox{6mm}{$7/8$} \\ \hline
\raisebox{6mm}9 & \includegraphics{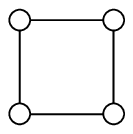} &  \raisebox{6mm}{$(3+\sqrt{3})/6$} 
&  \raisebox{6mm}{$(11+\sqrt{2})/14$} & \raisebox{6mm}{$(51+\sqrt{17})/76$} &  \raisebox{6mm}{$11/14$} \\ \hline
\raisebox{6mm}{10} & \includegraphics{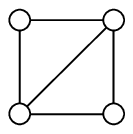} &  \raisebox{6mm}{$(9+\sqrt{3})/12$} 
&  \raisebox{6mm}{$(4+\sqrt{3})/6$} &  \raisebox{6mm}{$(3\sqrt{5}-5)/2$} &  \raisebox{6mm}{$8/9$}\\ \hline
%\raisebox{6mm}{11} & \includegraphics{k4.eps} &  \raisebox{6mm}{1} &  \raisebox{6mm}{1} &  \raisebox{6mm}{1} &  \raisebox{6mm}{1} \\ \hline
\end{tabular}
%\footnotetext{1- Simulation. 2- Exact formula in \eref{eq-alg}.}

\bigskip 

III. Trace monoids over 4 letters - exact values

\end{center}

\begin{center}
\begin{tabular}{|c|c|c|c|c|c|} \hline
&&&&& \\ 
& & $\lambda_{*}$ & $\lambda_{\M}$ &   $\gamma_{\M}^{-1}$ &  $\lambda_{\mathrm{cf}}$ \\
&&&&& \\ \hline\hline
\raisebox{6mm}1 & \includegraphics{i=k4.eps} &  \raisebox{6mm}{$0.25$} &  \raisebox{6mm}{$0.521\cdots$}   
&  \raisebox{6mm}{$0.4$}  & \raisebox{6mm}{1}  
\\ \hline
\raisebox{6mm}2 &\includegraphics{k2.eps} &  \raisebox{6mm}{0.5} &  \raisebox{6mm}{1}  
&  \raisebox{6mm}{$0.5$} & \raisebox{6mm}{1}  \\ \hline
\raisebox{6mm}3 &\includegraphics{k2k2.eps} &  \raisebox{6mm}{0.5} &  \raisebox{6mm}{0.75}  
&  \raisebox{6mm}{0.5} & \raisebox{6mm}{1}  \\ \hline
\raisebox{6mm}4 &\includegraphics{l3.eps} &  \raisebox{6mm}{$0.612\cdots$} 
&  \raisebox{6mm}{$0.923\cdots$} &  \raisebox{6mm}{$0.581\cdots$} &  
\;\; \raisebox{6mm}{$0.912\cdots$} \;\; \\ \hline
\raisebox{6mm}5 &\includegraphics{k3.eps} &  \raisebox{6mm}{0.75} &  \raisebox{6mm}{1}
&  \raisebox{6mm}{$0.667\cdots$} & \raisebox{6mm}{1}  \\ \hline
\raisebox{6mm}6 &\includegraphics{l4.eps} &  \raisebox{6mm}{$0.691\cdots$} & \raisebox{6mm}{$0.864\cdots$} 
&  \raisebox{6mm}{$0.643\cdots$} &  \raisebox{6mm}{$0.833\cdots$} \\ \hline
\raisebox{6mm}7 & \includegraphics{i=k3-4.eps} &  \raisebox{6mm}{$0.681\cdots$} & \raisebox{6mm}{$0.873\cdots$} 
&  \raisebox{6mm}{$0.676\cdots$} &  \raisebox{6mm}{$0.786\cdots$} \\ \hline
\raisebox{6mm}8 &\includegraphics{i=l2.eps} &  \raisebox{6mm}{$0.802\cdots$} 
&  \raisebox{6mm}{$0.927\cdots$} & \raisebox{6mm}{$0.760\cdots$} & \raisebox{6mm}{$0.875\cdots$} \\ \hline
\raisebox{6mm}{9} &\includegraphics{c4.eps} &  \raisebox{6mm}{$0.789\cdots$} 
&  \raisebox{6mm}{$0.887\cdots$} & \raisebox{6mm}{$0.725\cdots$} &  \raisebox{6mm}{$0.786\cdots$} \\ \hline
\raisebox{6mm}{10} &\includegraphics{i=k2.eps} &  \raisebox{6mm}{$0.894\cdots$} 
&  \raisebox{6mm}{$0.955\cdots$} &  \raisebox{6mm}{$0.854\cdots$} &  \raisebox{6mm}{$0.889\cdots$}\\ \hline
\end{tabular}

\bigskip 

III.b. Trace monoids over 4 letters - numerical values

\end{center}

Let us denote the dependence graphs in Table III, listed from top to bottom, by $(\Sigma,D_i), i=1,\dots, 10$. 
The graph $(\Sigma,D_9)$ is the cocktail party graph $CP_2$, hence the values of the 
average heights can be retrieved from section \ref{sse-hsg}. 
More generally, most of the values in the table can be computed using the results from the paper. 
The exceptions are 
$\lambda_{*}$ for $(\Sigma,D_i), i=6,7,8,$ and $10$. For $(\Sigma,D_8)$ and 
$(\Sigma,D_{10})$, the value of $\lambda_{*}$ can be computed by applying Proposition 12 from 
\cite{bril}. 

\medskip

For $(\Sigma,D_6)$, the exact value of $\lambda_*$ is not known. Using truncated Markov chains,
A. Jean-Marie \cite{ajm01} obtained the following exact bounds: 
\[
\lambda_{*}(\Sigma,D_6)\in [0.69125003165,0.69125003169]\:.
\] 

Let us concentrate on $\lambda_{*}(\Sigma,D_7)$.
Let $(x_n)_{n\in \N^*}$ be a sequence of independent 
random variables valued in $\Sigma$ and 
uniformly distributed: $P\{x_n=u\}=1/4, u\in \Sigma$. Define $X_n=\psi(x_1\cdots x_n)$, then $(X_n)_n$ is a 
Markov chain on the state space $\M(\Sigma,D_7)$. Let $a$ be the letter such that $(a,u)\in D_7$ 
for all $u\in \Sigma$. 
Define $T=\inf\{n \ : \ x_n=a\}$. An elementary argument using the Strong Law of Large Numbers then shows 
that $\lambda_{*}(\Sigma,D_7)=E[h(X_T)]/E[T]$. It follows that 
\begin{equation}\label{eq-d7}
\lambda_{*}(\Sigma,D_7)= \frac{1}{4} + \frac{1}{16} \left( \sum_{i\in \N} \frac{1}{4^i} \sum_{i_1+i_2+i_3=i}
\max(i_1,i_2,i_3) {i \choose i_1,i_2,i_3} \right).
\end{equation}
%This expression can be rewritten as follows
%\begin{equation}\label{eq-d7}
%\lambda_{*}(\Sigma,D_7) =
%{1 \over 4} + {1 \over 16}\, \left(\, \sum_{i=O}^{+\infty} \
%{(3\,i)! \over (i-1)!\, i!\, i!} \, {1 \over 64^i} \, \right)
%+ {3 \over 16}\, \left(\, \sum_{i>j,i\geq k} \
%{(i+j+k)! \over (i-1)!\, j!\, k!} \, {1 \over 4^{i+j+k}} \, \right) \:.
%\end{equation}
This expression involves non algebraic generalized hypergeometric
series. 
By truncating the infinite sum and upper-bounding the remainder using the inequality 
$\max(i_1,i_2,i_3)\leq i_1+i_2+i_3$, we get the following exact bounds:
\[
\lambda_{*}(\Sigma,D_7) \in [0.68111589347, 0.68111589349]\:.
\]
Another formula for $\lambda_{*}(\Sigma,D_7)$ involving multiple contour integrals 
and due to Alain Jean-Marie is given in \cite[Th. 13]{bril}. 

The closed form expressions for $\lambda_{\M}(\Sigma,D_7)$ and $\gamma_{\M}(\Sigma,D_7)$ 
are not given in Table III since they are too long and do not fit. We have
\begin{eqnarray}
\lambda_{\M}(\Sigma,D_7)& = & \frac{8(-93-9\sqrt{93}-\sqrt{93}X+5X^2)}{-1734-186\sqrt{93}+(141-5\sqrt{93})X+67X^2}
\nonumber \\
X & = & (108+12\sqrt{93})^{1/3}\:, \label{eq-l7}
\end{eqnarray}
and $\gamma_{\M}(\Sigma,D_7)^{-1}=$
\begin{equation}\label{eq-g7}
\frac{10777(529-23Y^2+Y^4)(829+132\sqrt{62}-(139-6\sqrt{62})Y-11Y^2)}
{3(3779+372\sqrt{62})(98340\sqrt{62}-1461365-1529(149+66\sqrt{62})Y-53885Y^2)}
\end{equation}
with $Y = (89+18\sqrt{62})^{1/3}$. 

\medskip

At last, let us comment on the value of $\gamma_{\M}$ for $(\Sigma,D_8)$. Using the 
results from section \ref{sse-upt}, we get
\begin{equation}\label{eq-alg}
\gamma_{\M}(\Sigma,D_8)^{-1}= \frac{(1-2\alpha)(4-5\alpha)}{7-27\alpha+24\alpha^2}\:,
\end{equation}
where $\alpha$ is the smallest root of the equation $2x^3-8x^2+6x-1=0$. Numerically, we have $\alpha=0.237\cdots$
and $\gamma_{\M}^{-1}=0.760\cdots$. 
In this case, Cardan's formulas are of no use (they provide an expression of the real
$\alpha$ as a function of the cubic root of a complex number). 

\medskip

Let us conclude by going back to the original motivation of compa\-ring the degree of parallelism 
in different trace monoids. We claim for instance that there is some strong evidence
that $(\Sigma,D_9)$ is `more parallel' than $(\Sigma,D_8)$. Indeed we have
$\lambda_*(\Sigma,D_9) < \lambda_*(\Sigma,D_8), \ \lambda_{\M}(\Sigma,D_9) < \lambda_{\M}(\Sigma,D_8), \ 
\gamma_{\M}^{-1}(\Sigma,D_9) < \gamma_{\M}^{-1}(\Sigma,D_8),$ and 
$\lambda_{\mathrm{cf}}(\Sigma,D_9) < \lambda_{\mathrm{cf}}(\Sigma,D_8)$.

\subsection*{Acknowledgement}

The authors would like to thank Mireille Bousquet-M\'elou and Xavier Viennot for pointing out several
relevant references. We are also grateful to Alain Jean-Marie for sharing with us his knowledge on the 
difficult problem of computing $\lambda_*$. 

%\bibliographystyle{plain}
%\bibliography{mairesse}
%\end{document}

\end{document}